**Title**

# Warming inhibits Increases in Vegetation Net Primary Productivity despite Greening in India

**Authors**

Ripan Das[1], Rajiv Kumar Chaturvedi[2], Adrija Roy[3], Subhankar Karmakar[1,4], Subimal Ghosh[1,3]

**Affiliations**

[1]Interdisciplinary Program in Climate Studies, Indian Institute of Technology Bombay, Powai, Mumbai - 400 076, India

[2] Department of Humanities and Social Sciences, Birla Institute of Technology and Science-Goa Campus, Zuarinagar, India

[3]Department of Civil Engineering, Indian Institute of Technology Bombay, Powai, Mumbai - 400 076, India

[4]Environmental Science and Engineering Department, Indian Institute of Technology Bombay, Powai, Mumbai - 400 076, India

[*]Corresponding Author. Email: subimal@iitb.ac.in

**Abstract**

India is the second-highest contributor to the post-2000 global greening. With satellite data, here we show that this 18.51% increase in Leaf Area Index (LAI) during 2001-2019 fails to translate into increased carbon uptake due to warming constraints. Our analysis further shows 6.19% decrease in Net Primary Productivity (NPP) during 2001-2019 over the temporally consistent forests in India despite 6.75% increase in LAI. We identify hotspots of statistically significant decreasing trends in NPP over the key forested regions of Northeast India, Peninsular India, and the Western Ghats. Together, these areas contribute to 31% of the NPP of India (1274.8 TgC.year $^{-1}$). These regions are the warming hotspots in India. Decreasing photosynthesis and stable respiration, above a threshold temperature, are the key reasons behind the declining NPP. Warming has already started affecting carbon uptake in Indian forests and calls for improved climate resilient forest management practices in a warming world.

**Introduction**

It is unequivocal that the anthropogenic Green House Gas (GHG) emissions are the key drivers of global warming since the pre-industrial period. Measurements show that during the last decade, the global average of annual anthropogenic $CO_2$ emissions has reached "the highest levels in human history" at $10.9 \pm 0.9$ PgC year$^{-1}$, 31% of which was stored by the terrestrial vegetation[1]. Terrestrial vegetation acts as a sink in the global carbon cycle through photosynthesis by uptaking the atmospheric $CO_2$ and recycling to its higher energy form, carbohydrate, and storing it as biomass[2]. Recent satellite observations have concluded that the Earth's green cover has increased significantly in the last two decades[3–6]. The 'greening' is generally measured with the Leaf Area Index (LAI), computed with one-sided leaf area for broadleaf canopies and half of the needle surface area for coniferous canopies[7]. The $CO_2$ fertilization effect[3,4] is the key contributor to the global greening trend. The other contributors include climatic factors[3,4,8] (example: the rising temperature in the high latitudes)  and the land use, land cover change[9,10].

This global greening intuitively enhances vegetation productivity and hence, increases the carbon sink potential. However, the recently published observed[11] and model-driven[12] studies have not found a proportional increase in vegetation productivity at a global scale. Gross Primary Productivity (GPP) is one of the key indicator for measuring vegetation productivity. GPP is defined as the amount of $CO_2$ captured by the plants in unit time during the time of photosynthesis. GPP has increased only 0.08 %, whereas the global green cover has increased about 0.23 % during the period 2000-2015. The lower increase in GPP is attributed to the atmospheric moisture stress globally[12]. Net Primary Productivity (NPP) is another important indicator for determining vegetation productivity.NPP is the difference between GPP and autotrophic respiration. There exists inconsistency in the global trends between NPP and GPP arising due to increasing autotrophic respiration[11,13]. Climate extremes like droughts and heatwaves are reported to be one of the key reasons behind the decline in terrestrial vegetation productivity [14–16].Temperature and precipitation are reported to be the two most important climatic factors that have control over vegetation productivity[8,16–18]. Zhang et al.[8] hypothesized that increasing temperature drives global NPP decrease, despite a stable GPP for the terrestrial ecosystem. Temperature[19], total water storage[20] and soil moisture[21] play major role in global carbon uptake by terrestrial vegetation and thus in the interannual variations of the Carbon Growth Rate (CGR).

Among the tropical regions, India is the second-highest contributor to post-2000 global greening after China[5]. However, it is not clear if such a greening resulted in an overall increase in primary productivity and thus in carbon uptake potential, given the impacts of changing climate on the vegetation. The majority of the studies also do not include the recent period and have considered pre-2010 years. Observational analysis with AVHRR data for the period 1982-2006 showed a 3.9% increasing trend of NPP/ decade in India, which has been attributed to $CO_2$ fertilization[22]. Other studies[23,24] with observational data and model output showed that precipitation is the major driver behind NPP variations. Nayak et al.[24] also estimated an increase of NPP at 0.005 PgC year$^{-2}$ during 1981-2006. Model-driven study[25] showed an increasing trend of NPP during 1901-2010 at 1.2-1.7 PgC.year$^{-2}$ due to elevated $CO_2$, LULC changes, and nitrogen deposition. A similar increasing trend of NPP was also reported for the period 2001-2006 with AVHRR and MODIS data[26]. One of the key limitations of the available studies is that NPP change analysis for the most recent decade is not yet reported in the literature. Here we use the MODIS quality-controlled satellite data of LAI, GPP, NPP, and Net Photosynthesis for the entire period of 2001-2019. We find a slight decreasing trend of NPP, and stable GPP during the 21$^{st}$ Century despite the increase in LAI. The results are contrary to the findings from earlier studies[22,24–26] that date back up to 2000s data. We further note that the biggest contributing regions of NPP in India have suffered from the larger decreasing trend. These are the regions in India, that experienced largest warming over 2001-2019. A statistical causality analysis confirmed that the decrease in NPP in India is attributable to the recent warming of 21$^{st}$ Century.

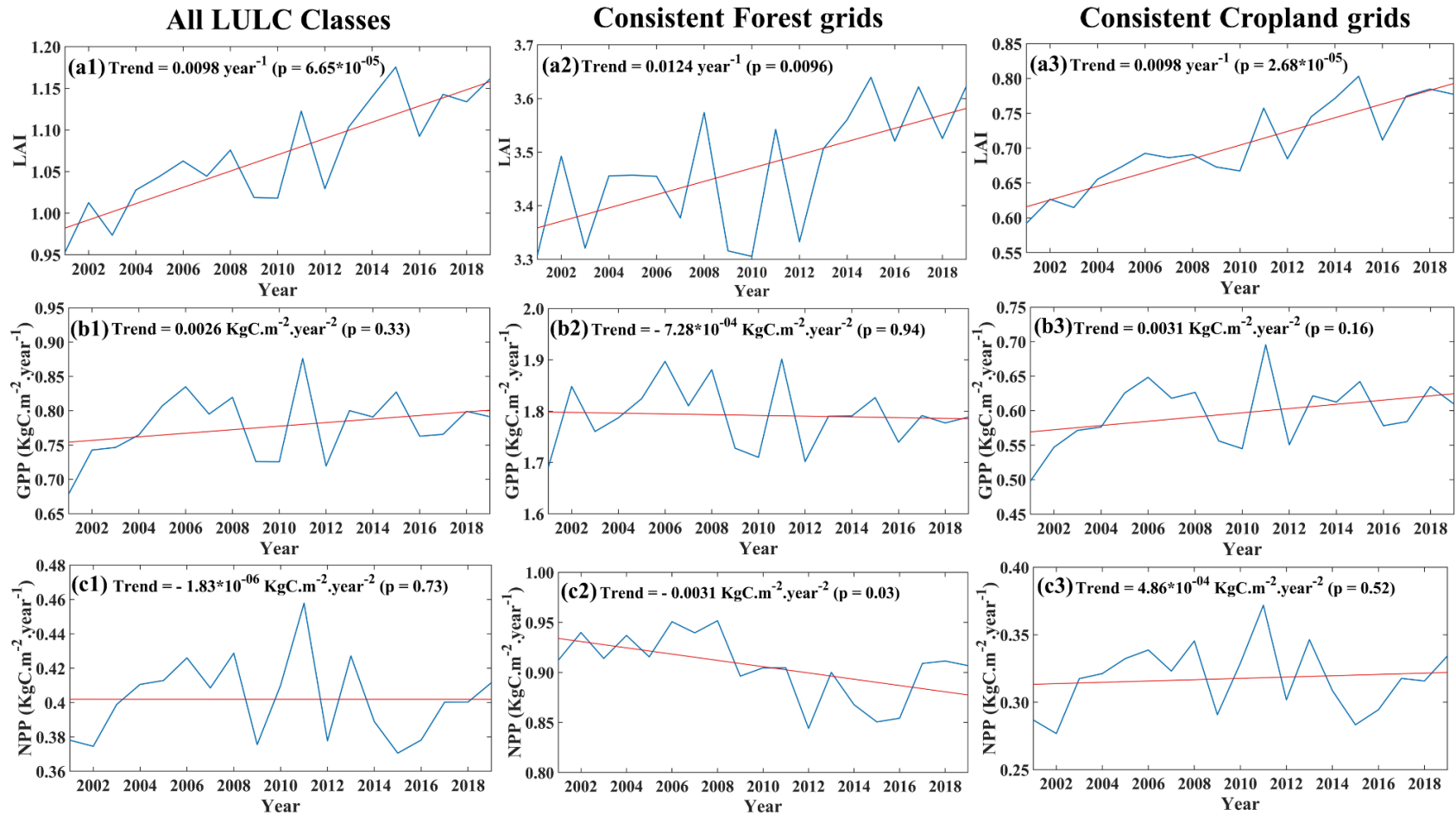

**Figure 1. Trend Inconsistencies between Greening and Carbon Uptake in India:** The time series of Leaf Area Index (a1), GPP (b1) and NPP (c1) over India. We have selected the grids over India where the land use was consistently forest and cropland during 2001-2019. The time-series of Leaf Area Index (a2), GPP (b2) and NPP (c2) over consistent forest grids in India. The time-series of Leaf Area Index (a3), GPP (b3) and NPP (c3) over consistent cropland grids in India.

**Results**

**Trends of LAI, GPP and NPP**

Figure 1 (a1-c1) presents the annual time series of LAI, GPP, and NPP over India. We find that the LAI over India has increased steadily (statistically significant) in the 21st century, which is consistent with the findings from Chen et al.[5], where they attributed the greening in India to the increased crop area. The annual time series of GPP and NPP have not shown any statistically significant trend despite this greening; the NPP showed a slight (not statistically significant) decreasing trend. MODIS NPP, GPP, and LAI  products are obtained by MODIS radiometric information. There is a high chance that these date sets are affected by covariability. Extensive ground truth verifications are also impossible for India due to the limited number of flux tower locations having long-term datasets. Furthermore, the MODIS products do not consider increasing productivity due to carbon fertilization [27]. Hence, we used two more datasets: FLUXCOM GPP and NEE products and Contiguous Solar-Induced chlorophyll Fluorescence (CSIF) data products for more robust analysis. Supplementary Figure 1 shows the climatologies of MODIS GPP, FLUXCOM GPP and CSIF over the Indian landmass. All three products show high photosynthesis during the last half of the monsoon, August and September. However, the values from MODIS maintain high GPP during the post-monsoon till February, while the other two climatologies drop significantly. It is noteworthy that none of these data products are observations and have their own limitations. It is expected that the GPP in India during post-monsoon should be maintained due to high radiations and high soil moisture from monsoon recharge. The same may continue in the winter, a cropping season in India known as Rabi season. MODIS GPP and CSIF dip during pre-monsoon summer because it is a dry, non-cropping season. The FLUXCOM GPP climatology shows an increase from February-March, which is unexpected. We plotted trends for CSIF, FLUXCOM GPP, and FLUXCOM NEE (reciprocal to net ecosystem carbon uptake) in the Supplementary Figure 2. The FLUXCOM GPP (Supplementary Figure 2b) shows a statistically non-significant decreasing trend, and FLUXCOM NEE (Supplementary Figure 2c) shows a statistically significant increasing trend over the Indian landmass. These are consistent with the results obtained from MODIS GPP and NPP trends. However, the CSIF (Supplementary Figure S2a) trends are the opposite of the other two datasets.

Upon investigation, we have found that CSIF is a machine learning-based extrapolated dataset for the study period (years 2001-2019) obtained based on the satellite datasets of SIF available from Orbiting Carbon Observatory-2 (OCO-2) for the period 2014-2017. The data may have limitations because it is an extrapolated dataset based on MODIS surface reflectance as the only predictor. The other two products showed that the greening in India has not been translated to the vegetation productivity.

Land use change (deforestation) [12] or climatic impacts [12,13] could have potentially inhibited the growth in GPP and NPP in India. The annual time series of both forests, and croplands show a steadily increasing trend (Supplementary Figure 3). Hence, land use change is apparently not the reason behind the non-increasing NPP at a country scale. We select the grid points, which were forestland consistently during the period 2001-2019 (Supplementary Figure 4), and present the time series of LAI, GPP, and NPP (Figure 1, a2, b2, and c2, respectively) spatially averaged over those grids. The LAI over the forestland shows a statistically significant increase. However, the GPP over the forestland shows a slight decrease (not significant), and the NPP shows a statistically significant decreasing trend. In India, forests are generally untouched by anthropogenic management interventions like irrigation, fertilizer use etc. One of the strong possibilities could be the impacts of changing climate on the forest ecosystem, which resulted in a decrease in primary productivity. Similar plots for the croplands show increased LAI (Figure 1 a3) a stable (no statistically significant trend) GPP and NPP (Figure 1, b3 and c3, respectively). The croplands are heavily managed, and hence, the natural impacts of climatic factors on croplands may be moderated by human interventions.

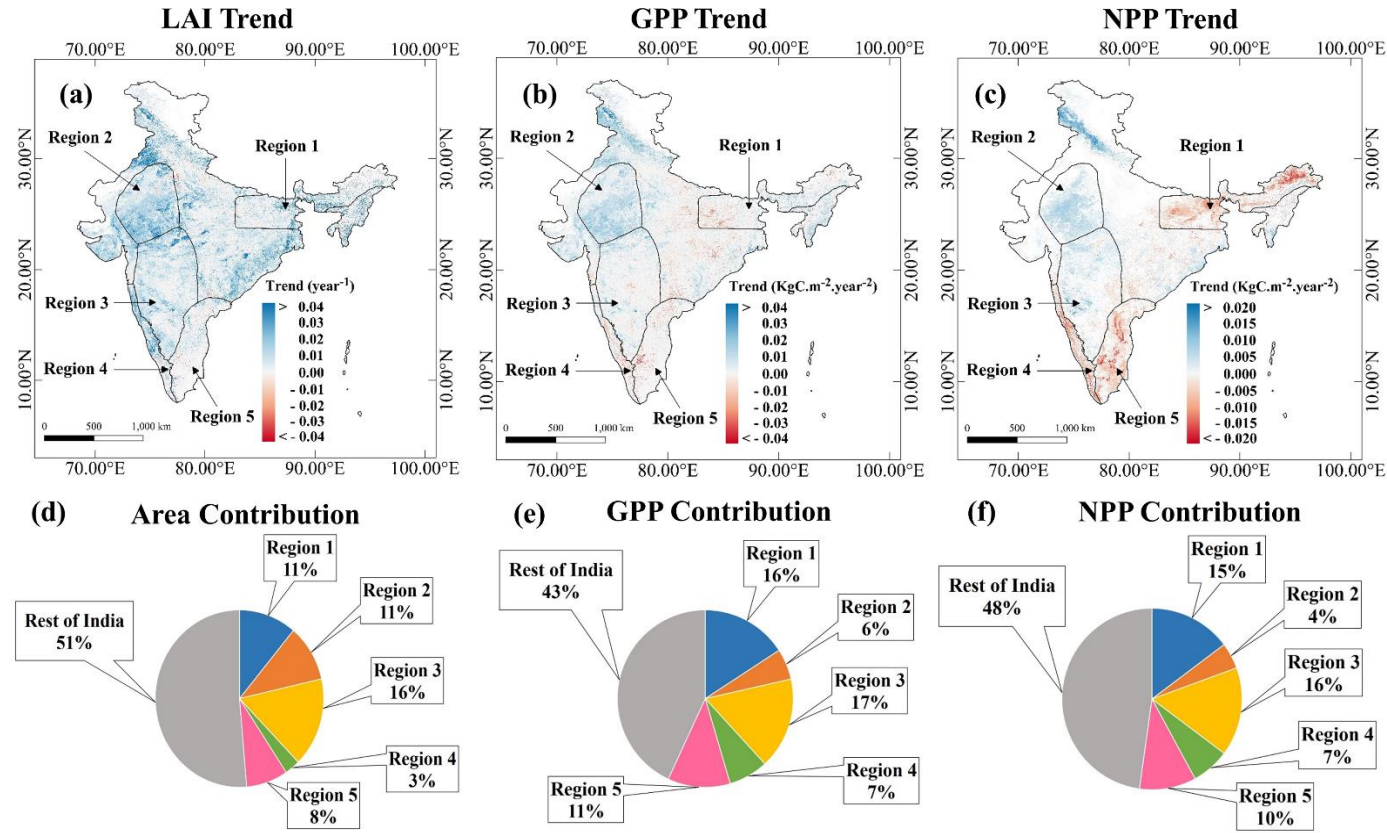

**Figure 2. Regional Trends:** Trends of LAI (a), GPP (b), and NPP (c) at statistically significant level 0.1. Based on the sign of trends, 5 regions were selected, Region 1: Northeast, Region 2: Northwest Arid, Region 3: Central Peninsular, Region 4: the Western Ghats, Region 5: East Coast Peninsular. Region 1, 4 and 5 have decreasing trends of NPP, while 2 and 3 have increasing trends of NPP. The area contributions (d) from regions 1, 4, and 5 to the total country area is 22%, whereas the GPP and NPP contributions are 34% and 32% of the national estimates, ((d) and (e)), respectively.

The spatial map of trends in MODIS LAI shows a widespread increase all over India (Figure 2a). However, the trends of GPP and NPP from the MODIS have strong spatial variations. Many of the regions over Northeast India and Peninsular India have a statistically significant decreasing trend of GPP, as evident from Figure 2b. The regions with decreasing NPP are more prominent over the Northeast and coastal peninsular India. Depending on the signs of the trends, we have selected 5 regions, Region 1: Northeast, Region 2: Northwest Arid, Region 3: Central Peninsular, Region 4: the Western Ghats, Region 5: East Coast Peninsular. Regions 1, 4, and 5 have statistically significant decreasing trends in NPP, whereas regions 2 and 3 have increasing NPP. We further standardized the MODIS LAI, MODIS GPP, and MODIS NPP datasets and calculated their trends. The standardization of these three variables place them on the same scale, and allows us to better understand their evaluations in different regions. The trend maps for the standardized MODIS LAI, MODIS GPP, and MODIS NPP datasets are shown in Supplementary Figure 5 (a, b, and c, respectively). The results of Supplementary Figure 5 (a-c) are consistent with those of Figure 2 (a-c). We have also plotted the spatial variations in the trends of CSIF, FLUXCOM GPP, and FLUXCOM NEE (Supplementary Figure 6, a, b, and c, respectively). We found strong increasing trends of NEE (signifying declining carbon uptake) almost everywhere in India. On the contrary, the trends of CSIF are increasing everywhere. The MODIS NPP trends are in between these two products. Apparently, the trends are contrasting among the products; however, there is a similarity. The CSIF product shows lower to no increasing trends only over regions 1, 4, and 5, while other areas in India have higher increasing trends. The MODIS NPP shows decreasing trends over regions 1, 4, and 5. The FLUXCOM shows high decreasing trends of carbon uptake over the same regions. These spatial patterns probably signify that vegetation carbon uptake over regions 1, 4, and 5 has been affected over the last two decades compared to the rest of India. Regions 1, 4 and 5 contribute to only 22% of India's area; however, their contribution to India's NPP and GPP are 34% and 32% (as per the MODIS product), respectively (Figure 2 d-f). It is noteworthy that the regions 1 and 4 contain the forests of Northeast India and the Western Ghats, which are among the world's eight hottest hotspots of biodiversity[28] and account for much of India's forest area and its forest carbon uptake. Now onward we are performing our analysis with the MODIS GPP and NPP, considering their spatial trend consistency with the other two products, FLUXCOM NEE (decreasing trend) and CSIF (slight increasing to no trend) in regions 1, 4, and 5.

We visually observed that in regions 1,4 and 5, the LAI trends diverge from NPP trend; to demonstrate this statistically, we generated a scatter plot of standardized LAI and NPP trends (Supplementary Figure 7). We also produced probability density functions (pdf) of standardized LAI and NPP trends using a bivariate kernel density estimate for each region (Supplementary Figure 8). Each point in the scatter plot and in the sample for generating pdfs represents grid. We kept the LAI (Standardized) trend on the x axis and the trend of NPP (Standardized) on the y axis. The density of LAI trend pixels is higher in the first and fourth quadrants in all five regions, signifying widespread greening over these five regions. We also found a trend divergence between LAI and NPP in regions 1 (Supplementary Figures 7a and 8a), 4 (Supplementary Figures 7d and 8d), and 5 (Supplementary Figures 7e and 8e). Pixel density (Supplementary Figure 8) was higher in the fourth quadrant of these three regions, where the LAI trend value is positive and the NPP trend value is negative. These findings are consistent with the trend maps (Figure 2 and Supplementary Figure 5). These plots allow us to demonstrate more quantitatively that LAI trend diverge from NPP trend during the study period in India's northeast (Region 1), western ghats (Region 4), and east coast peninsula (Region 5).

We zoom in the regions 1, 4, and 5, which have statistically significant decreasing trends in NPP, in Supplementary Figures 9, 10, and 11. For region 1, the Northeast India, the increase in the LAI is widespread (Supplementary Figure 9a) , though there are a few grids showing a decrease in LAI at the eastern side (Figure 2a). Such decreases are due to recent deforestation at isolated locations in the eastern states, such as Meghalaya, Arunachal Pradesh, and Nagaland[29]. However, the decreasing trends in NPP (Supplementary Figure 9c) are widespread over Northeast India, where there has been greening in most regions during 2001-2019. The GPP also has a decreasing trend except for central Northeast India. The Western Ghats region (Region-4) also has (Figure 2 and Supplementary Figure 10) decreasing GPP and NPP despite increasing LAI. Recent study[30] and the State of forest report in India[29] confirm that there is effectively no deforestation in the Western Ghats in the 21[st] Century. There are a few decreasing spots of LAI in the southern Western Ghats in the state of Kerala. This is probably due to declining paddy cultivation in Kerala that resulted from changing the agricultural practices to home gardening[31]. However, such isolated declining spots cannot result in a widespread decrease in NPP. In region 5, there are a few patches of declining LAI (Figure 2a) and in those patches, GPP and NPP have also declined (Figure 2 b and c) . However, like regions 1 and 4, the GPP and NPP decrease is widespread (Supplementary Figure 11, b and c). In Supplementary Figure 12 (a, c and e), we have plotted the time series of Forest cover area for region 1,4 and 5 respectively. We observed a steady increase in forest cover area in these regions. Cropland has also increased these regions (Supplementary Figure 12 b and f), except for region 4 (Supplementary Figure 12d). Therefore, land use and land cover changes cannot account for the divergence in trend between LAI and NPP over these regions.

We linearly regressed NPP as a function of LAI (NPP = f (LAI)) and also looked into the correlation between NPP and LAI to perform a more detailed analysis of the annual variation. In regions 1 (Supplementary Figure 13a) and 4 (Supplementary Figure 13d), we found that the goodness of fit ($R^2$) value was very low and statistically insignificant at level 0.1. Additionally, Supplementary Figure 14 shows a negative correlation between LAI and NPP over the majority of the regions in 1 and 4. Most regions of 2, 3, and 5 show a positive correlation (Supplementary Figure 14) between LAI and NPP. As a result, we can infer that the relationship between LAI and NPP is stronger in the regions (2 and 3) where there is no trend divergence between LAI and NPP. In contrast, the relationship between LAI and NPP is negative in the regions (1 and 4) where LAI and NPP trends diverge. Interestingly, the LAI over region 5 also does not show a strong positive trend or greening (Supplementary Figure 11a). Hence, the negative NPP does not result in a negative correlation in region 5, as observed in regions 1 and 4.

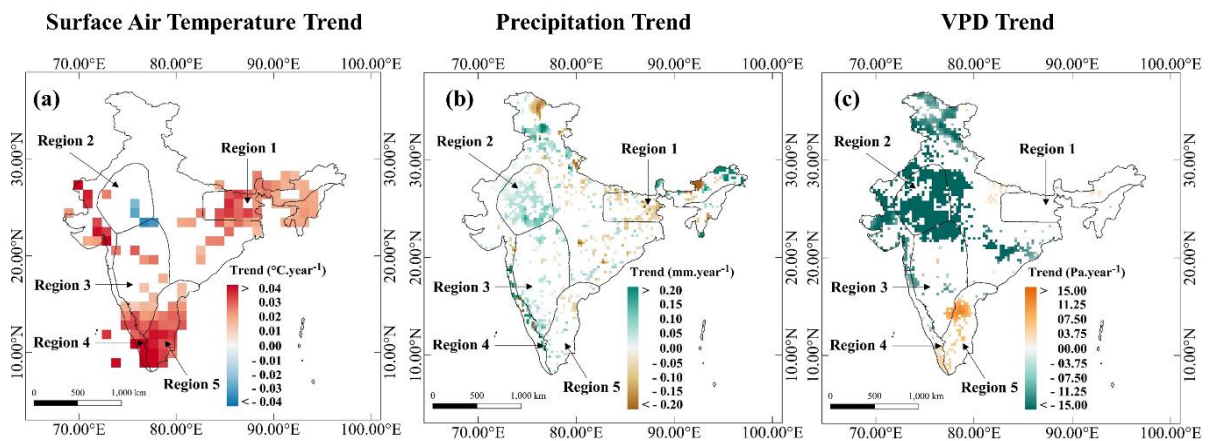

**Figure 3. Climatic Trends:** Trend of IMD Gridded Surface Air Temperature (a), IMD Gridded Precipitation (b), and Vapor Pressure Deficit (calculated using ERA5 reanalysis relative humidity and temperature data (c) in India for the period 2001-2019. The Figureures only show statistically significant trends at 0.1 level.

### Climate Controls on the Changing NPP/ GPP

To understand the climate connections to the declining NPP, we first present the trends of surface air temperature, precipitation, and vapor pressure deficit (Figure 3). warming is prominently visible (Figure 3a) only over regions 1, 4, and 5, pointing out a strong association with the declining GPP and NPP. Regions 1 and 5 experienced a mixed trend of precipitation (Figure 3b), whereas the precipitation has increased over region 4. There is a lack of similarity or spatial pattern match between the signs of precipitation trends and GPP/NPP trends. The Vapor Pressure Deficit (VPD) shows an increasing trend in some part over regions 4 and 5 (Figure 3c) that observed declining NPP/GPP. In the southern India, the VPD does not show significant changes despite warming (Figure 3a and Supplementary Figure 15a), which is probably because of probable increase in specific humidity (with increased moisture supply from ocean), with almost no change in relative humidity (Supplementary Figure 15b). VPD has decreased over region 2, where GPP/ NPP has a positive trend. There is no trend in VPD over region 1.

There is a spatial resemblance between India's warming hotspots and the regions with declining NPP, indicating a strong temperature control on the vegetation productivity. For a quantitiative understanding of the role of temperature on diverging trend of NPP and LAI, we linearly regressed NPP as a function of LAI (NPP = f (LAI)) and calculated the trend in the residuals for each grid (Supplementary Figure 16). The residual trend is decreasing in nature in regions (1, 4, and 5). In these regions (1,4 and 5), we also observed a significant warming trend (Figure 3a). We also generated the scatter plots between temperature and the residuals of linear regression (NPP against LAI, Supplementary Figure 17 (a,b,c,d,e)).We found that the unexplained (by LAI) components of NPP are negatively correlated with the temperature for all regions. These plots demonstrate the negative impacts of warming on the NPP.

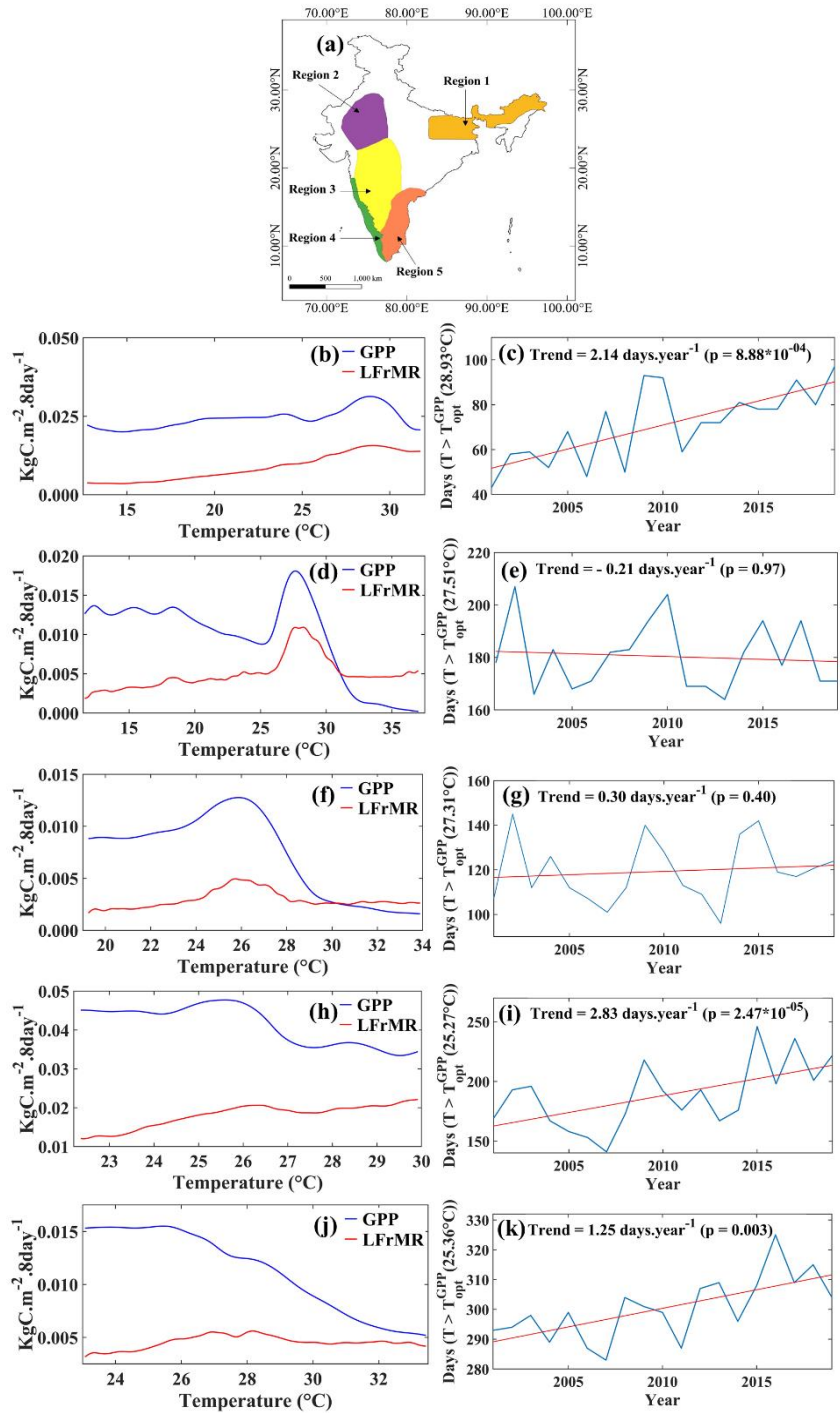

**Figure 4. Warming Impacts the Vegetation Productivity:** Map showing Regions (a), Sensitivity of Gross Primary Productivity (GPP) and Leaf and Fine root Maintenance respiration (LFrMR) to Temperature for Regions 1(b ), 2(d ), 3(f ), 4(h ), and 5 (j). The optimum temperature is the temperature at which GPP reaches the maximum. The GPP drops above this temperature. The trend of the number of days in a year exceeding optimum temperature for Regions 1(c ), 2(e ), 3(g ), 4(i ), and 5(k)

The GPP generally increases with temperature till an optimum value. When the temperature crosses the optimum value, GPP declines[32] while respiration rises[33], leading to decreasing NPP[34,35]. Tropical forests typically have the aforementioned characteristics, with a decrease of 9.1 megagrams of carbon uptake per hectare per degree Celsius warming in the mean daily maximum temperature in the warmest month[36]. We plot the variations of GPP in the 5 selected regions with temperature in Figure 4. The curves in the subplots are the smoothed variations as obtained using nonlinear kernel regression[37]. From these curves, we obtained the temperature value where GPP reaches the maximum, and from there, it starts dropping. That temperature is termed as optimum temperature. As MODIS provides only Leaf and Fine root Maintenance respiration (LFrMR) , we also used them in Figure 4. For regions 1, 4, and 5, the LFrMR are either stable or increase beyond the optimum temperature. Hence, with warming above the optimum temperature, the net Photosynthesis (PSNnet, the difference between GPP and LFrMR) should decrease in these regions with a subsequent decline of NPP. We also investigated the equations used to calculate the MODIS NPP and solved them analytically (Supplementary Note 1) to looked into the temperature response leaf and fine root Maintenance respiration rates individually and collectively (Supplementary Figure 18). We found that the rate of change in Leaf Maintenance Respiration (Leaf MR) with temperature drops after 25° C (Supplementary Figure 18a); however, it remains positive, suggesting an increase in leaf Maintenance respiration with the temperature. The rate of change in fine root Maintenance respiration keeps on increasing with temperature (Supplementary Figure 18b) and the total rate of change in Leaf and Fine root Maintenance respiration becomes stable after 30° C (Supplementary Figure 18c). The results are very similar to those obtained from Figure. 4 (b, d, f, h and j) showing the consistency.The time series and the trend of the number of days in a year above the optimum temperature ($T^{GPP}_{opt}$) for the regions are presented in Figure. 4 c, e, g,i and k . For regions 1, 4, and 5, there are statistically significant increasing trends in the number of days above optimum temperature ($T^{GPP}_{opt}$). These trends, along with low photosynthesis above optimum temperature, results in low net carbon uptake by the vegetation. Interestingly, for regions 2 and 3, the LFrMR drops after the optimum temperature; however, the drops are not as high as the drops in GPP. Region 2 is India's warmest and arid region, and probably the vegetation is already adapted to a warm environment. The decrease in LFrMR could be because of this long-term adaptation of vegetation. For regions 2 and 3, the numbers of days exceeding the optimum temperature

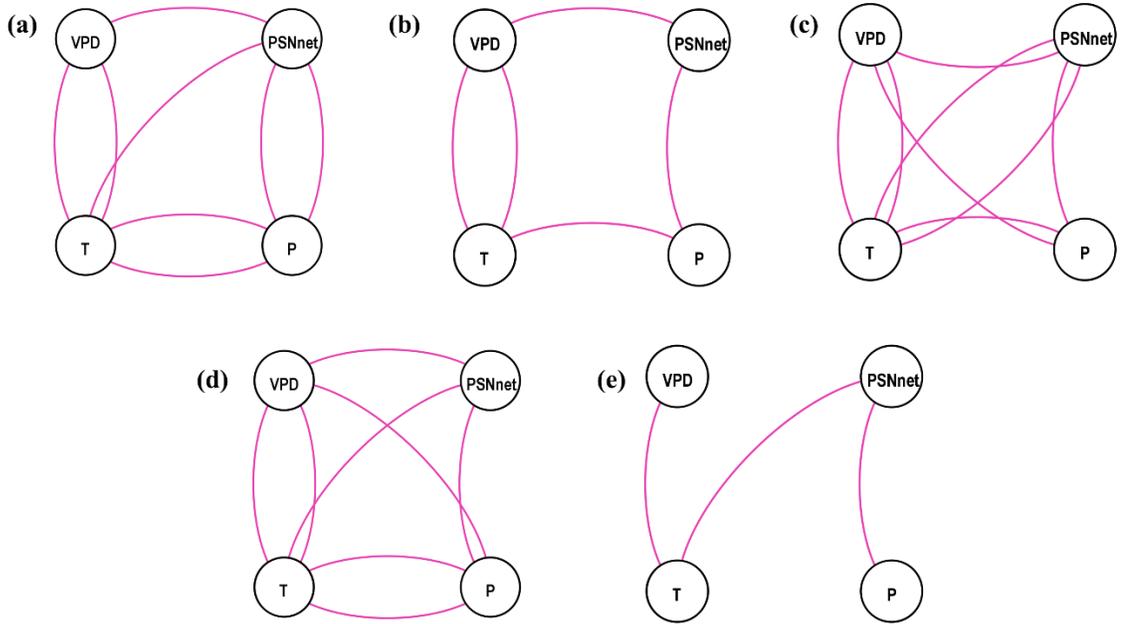

do not have a trend. Hence, the decline in NPP is not observed in these regions.

**Figure 5. Causal Network between Climate and Productivity Variables:** The causal network derived from Granger Causality at the regions 1 (a), 2(b), 3(c), 4(d), 5 (e). The variables used are Precipitation (P), Temperature (T), Vapor Pressure Deficit (VPD), and net Photosynthesis (PSNnet). The causal links are from source to sink in a clockwise direction at statistically significance level 0.05.

We use the Granger causality approach to further ascertain the climate controls on changing PSNnet (and subsequently, NPP). The results are presented in Figure. 5 for all the 5 regions. Except for region 2, for all the other regions, there are causal links from temperature to PSNnet. For region 2, temperature causes vapor pressure deficit that subsequently causes PSNnet. Hence, for all the regions, the causal links from temperature to PSNnet are established. Figure. 2-5 collectively show that the warming impacts the net carbon uptake by vegetation with a declining trend of NPP over regions 1, 4, and 5. It should be noted that precipitation has a causal connection to PSNnet as expected over the monsoon region of India. However, there is no consistent trend similarity between Precipitation and GPP or NPP. Hence, the causal link from Precipitation to PSNnet in Figure. 5 presents the covariation of the variables at seasonal or intra-seasonal scale as reported by Valsala et al.[23]. To further confirm the causal connections, we plot the causal networks with GPP in Supplementary Figure 19, similar to Figure. 5. They show that for all the regions, the temperature has causal connections either to GPP hence affects the PSNnet and NPP. Using Granger Causality, we may identify the causal factors; however, quantifying contributions from a causal variable to an impact variable is difficult. Furthermore, all the meteorological variables are connected, with the possibility of multiple confounding factor(s). Hence, we did not attempt to quantify the exact causal contribution from temperature to productivity. This is one of the limitation of the present causal analysis and probably a generalized limitation of any widely used causal discovery methods.

Carbon fertilization plays a role in improving water use efficiency and primary productivity and shoule be considered as a potential causal factor in Granger Causality analysis. However, we do not have any available long-term dataset in India that considers carbon fertilization. This is a major limitation of the study. Further, the atmospheric CO2 concentration does not change with a high-frequency variability (monthly scale) that was used in the Granger Causality analysis. Considering annual CO2 will result in a small annual sample insufficient to perform any causal analysis.

**Discussion**

Recent decades have experienced increasing atmospheric $CO_2$ concentration and widespread warming around the globe. India is not an exception. Increased warming resulted in an increase in water vapor carrying capacity and subsequent increase in the vapor pressure deficit. Stomatal conductance reduces under increased VPD and transpiration increases till a VPD threshold [38,39]. Both of these processes lead to reduced photosynthesis and these impacts have been observed globally [38,39] Global studies showed that the VPD increased by $0.0017 \pm 0.0001$ kPa year−1 (Data: CRU) the GPP decreased by $0.23 \pm 0.09$ Pg C year−1 and $0.31 \pm 0.11$ Pg C year−1 since the 1999, as per the EC-LUE and MODIS models, respectively[40]. On the contrary, model-driven studies show improved WUE [41–43] and vegetation productivity under increased $CO_2$ concentration due to carbon fertilization [41,42,44]. The science question remains if Carbon fertilization can make up for the reduced productivity of the vegetation due to warming and the increased VPD. A synthesis study performed in the IPCC Assessment Report 6, Working Group 1, Chapter 5 [1] and Chapter 11 [45] showed with a high confidence that the vegetation carbon sink will be less efficient at a very high warming level. Major limitations in such studies are: 1) the models are still inadequate to consider the physiological response of plants to the increased VPD, and 2) the satellite products like MODIS do not consider Carbon fertilization. The opposing impacts of both the factors and the resulting uncertainty in projections are more pronounced at a lower warming level. The literature agrees that high VPD impacts will dominate globally at a higher warming level over Carbon fertilization due to multiple direct and indirect effects [1].

The lack of ground-truth data availability in India is one of the major bottlenecks in performing a detailed analysis of vegetation productivity trends and responses to increased VPD and carbon fertilization. The primary contributor to greening in India is agricultural expansion [5]. Recent studies show decreased crop yield due to warming [46]. This finding is probably one of the proxies that confirm the warming-induced declining vegetation productivity in India and is in agreement with our conclusions from the present analysis.

In the last two decades, the green cover of India has increased [MoEFCC, 2021; https://unfccc.int/sites/default/files/resource/INDIA_%20BUR-3_20.02.2021_High.pdf"] and Indian forests are still a net sink of carbon[47]; however, our study finds that this sink may be weakening due to warming. The NPP from temporally consistent Indian forests shows a significant declining trend of **-0.75±0.58 Tg C year$^{-2}$** and has declined by 6.19% over the 19 years in the 21$^{st}$ century. The most prominent vegetation productivity decline has been witnessed in some of India's most biodiverse and pristine forest regions such as Northeast India and Western Ghats. These regions are also found to be the warming hotspots in the country. Variations of GPP and net photosynthesis with temperature and causal analysis prove that warming has reduced the vegetation productivity in India despite greening. Our analysis shows that climate change especially Temperature rise has already started affecting vegetation productivity and carbon uptake in Indian forests. The forest sinks make for a significant part of India's climate mitigation pledges till 2030

(https://www4.unfccc.int/sites/ndcstaging/PublishedDocuments/India%20First/INDIA%20INDC%20TO%20UNFCCC.pdf), and possible net zero plans till 2070. Hence, our Our results have significant implications on the country's forest conservation, biodiversity conservation as well as its carbon sink plans including India's possible net zero plans till 2070.

## Methods

**MODIS LAI product:** Terra MODIS (MOD15A2H) Version 6 datasets are used to analyze vegetation greenness[7]. The spatial and temporal resolutions are 500 meters (m) and 8 days, respectively. The data quality control layer obtained from the data source is applied to remove the cloud-covered and bad pixels. In the Supplementary Figure 20, we calculated the annual average percentage of good pixels taken from each grid. Annually, 82% to 89% were the good samples per grid, averaged across the grids, considered during the study period. We also presented a spatial map of overall good pixel count at the grid level in the Supplementary Figure 21. We discovered that, on average, more than 85% of the pixels in each grid were taken from 2001 to 2019. Except for a very few grids, almost all the grids have more than 70% good pixels. Since good pixel count never reaches low (<50%) for any grids, we have not performed any sensitivity analysis. The data is processed as per the user guidelines[7]. These 8-day data sets of LAI are converted into monthly and annual datasets.

**MODIS GPP and PSNnet Product:** In this study, we used MOD17A2HGF Version 6 Gross Primary Productivity (GPP) product and Net photosynthesis (PSNnet) product, with the spatial and temporal resolutions are 500 meters (m) and 8 days respectively, for calculating vegetation productivity. MOD17A2HGF Version 6 achieved stage 3 validation[48], this current version has not validate globally but previous versions of MOD17A2H GPP datasets are validate globally across different biomes with Fluextower GPP in many studies[49,50] ($R^2 > 0.50$). Net photosynthesis (PSNnet) is the difference between GPP and leaf and root respiration. We used the 8-day cumulative GPP and PSNnet product. These datasets also contain a Data quality control layer which we applied for removing the cloud-covered and bad pixels. We have not performed any sensitivity analysis as mentioned earlier.We processed the data as per the user guidelines[48] and converted these 8-day data sets of GPP and PSNnet into monthly and annual datasets.

**MODIS NPP Product:** We obtained the Net Primary productivity from MOD17A3HGF Version 6 product; it gives us annual net primary productivity at 500 meters (m) spatial resolution. Previous version of MOD17A3H dataset are consistent with Fluxtower NPP reflecting $R^2$ value of 0.56[51]. Annual NPP is calculated by taking the difference between GPP and Autotrophic respiration (RA) which is the sum of Growth respiration (RG) and Maintenance respiration (RM). This data is processed as per the user guidelines[52].

**CSIF Product**: CSIF or contiguous solar-induced chlorophyll fluorescence data is generated by training a neural network (NN) with surface reflectance from the MODIS and SIF from Orbiting Carbon Observatory-2 (OCO-2). It was trained for the year 2015 and 2016 and validate it for 2014 and 2017. The data used in the study are from clear sky condition. The spatial resolution of these data set is $0.5° \times 0.5°$ and the temporal resolution of this data set is 4 days [53]

**FLUXCOM GPP and NEE Product :** FLUXCOM GPP and NEE products are generated by 3 machine learning methods (RF, ANN, MARS) with the help of Remote sensing data from MODIS (LST, NDVI, EVI, LAI, fPAR, BRDF) and Meteorological data from CRUNCEPv6 (Air Temperature, Precipitation, Global radiation and VPD). This dataset is available at Monthly scale at $0.5° \times 0.5°$ resolution. The datasets are further evaluated with the eddy covariance data, for GPP $R^2 > 0.7$ and for NEE $R^2 < 0.5$. [54,55]

**MODIS LULC Product :** Changes in Forest cover and Croplands are calculated using MODIS MCD12Q1 datasets. This dataset gives us an annual land cover map at 500 meters (m) spatial resolution globally with different land cover legends. In this study, we used IGBP classification to detect the land cover type. Evergreen Needleleaf Forests, Evergreen Broadleaf Forests, Deciduous Needleleaf Forests, Deciduous Broadleaf Forests, and Mixed Forests are considered as the forest cover. For Cropland, we considered the grids with 60% area as cultivated[56].

**Precipitation and Temperature Data :** Daily gridded precipitation data at $0.25° \times 0.25°$ resolution provided by India Meteorological Department (IMD) is used in this study[57]. Daily average surface temperature is also obtained from IMD at $1° \times 1°$ resolution[58] Monthly average and annual average precipitation and temperature are then obtained from these daily datasets.

**Vapour Pressure Deficit (VPD) Calculation :** Vapour Pressure Deficit is the difference between saturation water vapour pressure and actual water vapour pressure. We calculated VPD by using Teten formula (Equation 1) [59] in SI unit

$$VPD = 611 \times e^{\left(\frac{17.27 \times t}{t + 237.3}\right)} \times \left(1 - \left(\frac{RH}{100}\right)\right) \tag{1}$$

Here t is the average temperature and RH is the relative humidity in equation 1. We used European Centre for Medium-Range Weather Forecasts (ECMWF) ERA5 monthly averaged reanalysis datasets of Relative humidity and temperature at $0.25° \times 0.25°$ spatial resolution at 1000 hPa level[60] . Monthly averaged VPD are then converted into annual averaged VPD.

**Calculation of Trend:** We used the modified Mann-Kendall test[61] for detecting the trend. We use the significance level $p \leq 0.10$ if not mentioned. After the detection of the trend, the slopes of the trend lines are obtained using linear regression. The percentage changes for 2001-2019 are calculated based on trend, duration and initial value at 2001. They are computed by multiplying the annual statistical significant trend with the duration (18 years), divided by the initial value at the starting year, 2001.

**Granger causality test**: Granger causality test is a statistical hypothesis test that helps to find the cause-effect relationship between two time series. We say that 'X' is causing 'Y', only if we get a better prediction of 'Y' by including the past values of 'X' in the predictor[62] . In this study, we used the Granger causality test between the monthly datasets of Temperature (T), Rainfall (R), Vapour Pressure deficit (VPD), and Net photosynthesis (PSNnet) / Gross primary productivity (GPP) for finding the causal relationships among them. We detrended each time series before applying the Granger Causality. We have considered maximum lag up to 11 months and significance level p ≤ 0.05. The detrended time series are put into a vector autoregression (VAR) model and then "leave-one-out" type granger causality test is performed. The null hypothesis is tested by conducting "chi-square test".

**Data availability:** MODIS LAI data at 8 day, 500 m spatial resolution has been used in this study. MOD15A2H data can be downloaded from https://ladsweb.modaps.eosdis.nasa.gov/search/order/2/MOD15A2H--6 . MODIS GPP and PSNnet data at 8 day, 500 m spatial resolution data can be downloaded from https://ladsweb.modaps.eosdis.nasa.gov/search/order/2/MOD17A2HGF--6 . MODIS Annual NPP data at 500 m spatial resolution data can be downloaded from https://ladsweb.modaps.eosdis.nasa.gov/search/order/2/MOD17A3HGF--6. CSIF data at 0.5° spatial resolution and 4 day temporal resolution can be downloaded from https://Figureshare.com/articles/dataset/CSIF/6387494. Monthly FLUXCOM GPP and NEE datasets at at 0.5°× 0.5° spatial resolution can be downloaded from https://www.bgc-jena.mpg.de/geodb/projects/Data.php. MODIS LULC data at 500 m spatial resolution can be downloaded from https://ladsweb.modaps.eosdis.nasa.gov/search/order/2/MCD12Q1--6 . India Meteorological Department (IMD) provided gridded daily rainfall data at 0.25°× 0.25° spatial resolution and gridded daily maximum temperature and minimum temperature at 1°×1° spatial resolution. IMD gridded Rainfall and Temperature data can be downloaded from https://imdpune.gov.in/Clim_Pred_LRF_New/Grided_Data_Download.html. ECMWF reanalysis dataset of Monthly Relative Humidity and monthly mean temperature data at 0.25°× 0.25° spatial resolution are downloaded from https://cds.climate.copernicus.eu/cdsapp#!/dataset/reanalysis-era5-pressure-levels-monthly-means?tab=form to calculate VPD.

**References:**

1.    Canadell, J. G. *et al.* Global Carbon and other Biogeochemical Cycles and Feedbacks
Supplementary Material. in *Climate Change 2021: The Physical Science Basis.*
*Contribution of Working Group I to the Sixth Assessment Report of the Intergovernmental*
*Panel on Climate Change* (2021).

2.    Fatichi, S., Pappas, C., Zscheischler, J. & Leuzinger, S. Modelling carbon sources and
sinks in terrestrial vegetation. *New Phytol.* **221**, 652–668 (2019).

3.    Zhu, Z. *et al.* Greening of the Earth and its drivers. *Nat. Clim. Chang.* **6**, 791–795 (2016).

4.    Piao, S. *et al.* Characteristics, drivers and feedbacks of global greening. *Nat. Rev. Earth*
*Environ.* **1**, 14–27 (2020).

5.    Chen, C. *et al.* China and India lead in greening of the world through land-use
management. *Nat. Sustain.* **2**, 122–129 (2019).

6.    Cortés, J. *et al.* Where Are Global Vegetation Greening and Browning Trends Significant?
*Geophys. Res. Lett.* **48**, (2021).

7.    Myneni, R., Knyazikhin, Y. & Park, T. *MYD15A2H MODIS/Aqua Leaf Area Index/FPAR*
*8-Day L4 Global 500m SIN Grid V006*. (2015)
doi:https://doi.org/10.5067/MODIS/MYD15A2H.006.

8.    Zhang, Y. *et al.* Climate-driven global changes in carbon use efficiency. *Glob. Ecol.*
*Biogeogr.* **23**, 144–155 (2014).

9.    Tong, X. *et al.* Increased vegetation growth and carbon stock in China karst via ecological
engineering. *Nat. Sustain.* **1**, 44–50 (2018).

10.   Mao, J. *et al.* Human-induced greening of the northern extratropical land surface. *Nat.*
*Clim. Chang.* **6**, 959–963 (2016).

11.   Ding, Z., Peng, J., Qiu, S. & Zhao, Y. Nearly Half of Global Vegetated Area Experienced

Inconsistent Vegetation Growth in Terms of Greenness, Cover, and Productivity. *Earth's Futur.* **8**, (2020).

12. Zhang, Y., Song, C., Band, L. E. & Sun, G. No Proportional Increase of Terrestrial Gross Carbon Sequestration From the Greening Earth. *J. Geophys. Res. Biogeosciences* **124**, 2540–2553 (2019).

13. Wu, L. *et al.* Climate change weakens the positive effect of human activities on karst vegetation productivity restoration in southern China. *Ecol. Indic.* **115**, 106392 (2020).

14. Zhao, M. & Running, S. W. Drought-Induced Reduction in Global Terrestrial Net Primary Production from 2000 Through 2009. *Science (80-. ).* **329**, 940–943 (2010).

15. Ciais, P. *et al.* Europe-wide reduction in primary productivity caused by the heat and drought in 2003. *Nature* **437**, 529–533 (2005).

16. Reichstein, M. *et al.* Climate extremes and the carbon cycle. *Nature* **500**, 287–295 (2013).

17. He, Y., Piao, S., Li, X., Chen, A. & Qin, D. Global patterns of vegetation carbon use efficiency and their climate drivers deduced from MODIS satellite data and process-based models. *Agric. For. Meteorol.* **256**–**257**, 150–158 (2018).

18. Pan, S. *et al.* Impacts of climate variability and extremes on global net primary production in the first decade of the 21st century. *J. Geogr. Sci.* **25**, 1027–1044 (2015).

19. Wang, X. *et al.* A two-fold increase of carbon cycle sensitivity to tropical temperature variations. *Nature* **506**, 212–215 (2014).

20. Humphrey, V. *et al.* Sensitivity of atmospheric CO2 growth rate to observed changes in terrestrial water storage. *Nature* **560**, 628–631 (2018).

21. Humphrey, V. *et al.* Soil moisture–atmosphere feedback dominates land carbon uptake variability. *Nature* **592**, 65–69 (2021).

22. Bala, G. *et al.* Trends and Variability of AVHRR-Derived NPP in India. *Remote Sens.* **5**,

810–829 (2013).

23. Valsala, V. *et al.* Intraseasonal variability of terrestrial biospheric CO 2 fluxes over India during summer monsoons. *J. Geophys. Res. Biogeosciences* **118**, 752–769 (2013).

24. Nayak, R. K., Patel, N. R. & Dadhwal, V. K. Inter-annual variability and climate control of terrestrial net primary productivity over India. *Int. J. Climatol.* **33**, 132–142 (2013).

25. Banger, K. *et al.* Terrestrial net primary productivity in India during 1901–2010: contributions from multiple environmental changes. *Clim. Change* **132**, 575–588 (2015).

26. Nayak, R. K. *et al.* Assessing the consistency between AVHRR and MODIS NDVI datasets for estimating terrestrial net primary productivity over India. *J. Earth Syst. Sci.* **125**, 1189–1204 (2016).

27. Keenan, T. F. *et al.* RETRACTED ARTICLE: A constraint on historic growth in global photosynthesis due to increasing CO2. *Nature* **600**, 253–258 (2021).

28. Myers, N., Mittermeier, R. A., Mittermeier, C. G., da Fonseca, G. A. B. & Kent, J. Biodiversity hotspots for conservation priorities. *Nature* **403**, 853–858 (2000).

29. Forest Survey of India. *Forest Cover: India State of Forest Report 2019 Volume 1 (Chapter 2)*. https://fsi.nic.in/isfr19/vol1/chapter2.pdf (2019).

30. REDDY, C. S., JHA, C. S. & DADHWAL, V. K. Assessment and monitoring of long-term forest cover changes (1920–2013) in Western Ghats biodiversity hotspot. *J. Earth Syst. Sci.* **125**, 103–114 (2016).

31. Fox, T. A. *et al.* Agricultural land-use change in Kerala, India: Perspectives from above and below the canopy. *Agric. Ecosyst. Environ.* **245**, 1–10 (2017).

32. Moore, C. E. *et al.* The effect of increasing temperature on crop photosynthesis: from enzymes to ecosystems. *J. Exp. Bot.* **72**, 2822–2844 (2021).

33. Heskel, M. A. *et al.* Convergence in the temperature response of leaf respiration across

biomes and plant functional types. *Proc. Natl. Acad. Sci.* **113**, 3832–3837 (2016).

34. Dusenge, M. E., Duarte, A. G. & Way, D. A. Plant carbon metabolism and climate change: elevated CO 2 and temperature impacts on photosynthesis, photorespiration and respiration. *New Phytol.* **221**, 32–49 (2019).

35. Niu, S. *et al.* Thermal optimality of net ecosystem exchange of carbon dioxide and underlying mechanisms. *New Phytol.* **194**, 775–783 (2012).

36. Sullivan, M. J. P. *et al.* Long-term thermal sensitivity of Earth's tropical forests. *Science (80-. ).* **368**, 869–874 (2020).

37. Yi Cao. Kernel Smoothing Regression. (2021).

38. Grossiord, C. *et al.* Plant responses to rising vapor pressure deficit. *New Phytol.* **226**, 1550–1566 (2020).

39. Breshears, D. D. *et al.* The critical amplifying role of increasing atmospheric moisture demand on tree mortality and associated regional die-off. *Front. Plant Sci.* **4**, (2013).

40. Yuan, W. *et al.* Increased atmospheric vapor pressure deficit reduces global vegetation growth. *Sci. Adv.* **5**, (2019).

41. Keenan, T. F. *et al.* Increase in forest water-use efficiency as atmospheric carbon dioxide concentrations rise. *Nature* **499**, 324–327 (2013).

42. El Masri, B. *et al.* Carbon and Water Use Efficiencies: A Comparative Analysis of Ten Terrestrial Ecosystem Models under Changing Climate. *Sci. Rep.* **9**, 14680 (2019).

43. Warren, J. M. *et al.* Ecohydrologic impact of reduced stomatal conductance in forests exposed to elevated CO2. *Ecohydrology* **4**, 196–210 (2011).

44. Norby, R. J. & Zak, D. R. Ecological Lessons from Free-Air CO 2 Enrichment (FACE) Experiments. *Annu. Rev. Ecol. Evol. Syst.* **42**, 181–203 (2011).

45.     Seneviratne, S. I. *et al.* 11 Chapter 11: Weather and climate extreme events in a changing climate. (2021) doi:10.1017/9781009157896.013.

46.     Lobell, D. B., Schlenker, W. & Costa-Roberts, J. Climate Trends and Global Crop Production Since 1980. *Science (80-. ).* **333**, 616–620 (2011).

47.     Harris, N. L. *et al.* Global maps of twenty-first century forest carbon fluxes. *Nat. Clim. Chang.* **11**, 234–240 (2021).

48.     Running, S., Mu, Q., Zhao, M. & Moreno, A. MOD17A2HGF MODIS/Terra Gross Primary Productivity Gap-Filled 8-Day L4 Global 500 m SIN Grid V006 [Data set]. NASA EOSDIS Land Processes DAAC. 1–38 (2019) doi:https://doi.org/10.5067/MODIS/MOD17A2HGF.006.

49.     Wang, L. *et al.* Evaluation of the Latest MODIS GPP Products across Multiple Biomes Using Global Eddy Covariance Flux Data. *Remote Sens.* **9**, 418 (2017).

50.     Tang, X. *et al.* A comprehensive assessment of MODIS-derived GPP for forest ecosystems using the site-level FLUXNET database. *Environ. Earth Sci.* **74**, 5907–5918 (2015).

51.     Peng, D. *et al.* Country-level net primary production distribution and response to drought and land cover change. *Sci. Total Environ.* **574**, 65–77 (2017).

52.     Running, S., Mu, Q., Zhao, M. & Moreno, A. MOD17A3HGF MODIS/Terra Net Primary Production Gap-Filled Yearly L4 Global 500 m SIN Grid V006 [Data set]. NASA EOSDIS Land Processes DAAC. (2019) doi:https://doi.org/10.5067/MODIS/MOD17A3HGF.006.

53.     Zhang, Y., Joiner, J., Alemohammad, S. H., Zhou, S. & Gentine, P. A global spatially contiguous solar-induced fluorescence (CSIF) dataset using neural networks. *Biogeosciences* **15**, 5779–5800 (2018).

54.     Tramontana, G. *et al.* Predicting carbon dioxide and energy fluxes across global

FLUXNET sites with regression algorithms. *Biogeosciences* **13**, 4291–4313 (2016).

55.    Jung, M. *et al.* Compensatory water effects link yearly global land CO2 sink changes to temperature. *Nature* **541**, 516–520 (2017).

56.    Gray, J., Sulla-Menashe, D. & Friedl, M. A. User Guide to Collection 6 MODIS Land Cover Dynamics (MCD12Q2) Product. *User Guid.* **6**, 1–8 (2019).

57.    Pai, D. S. *et al.* Development of a new high spatial resolution (0.25° × 0.25°) long period (1901-2010) daily gridded rainfall data set over India and its comparison with existing data sets over the region. *Mausam* **65**, 1–18 (2014).

58.    Srivastava, A. K., Rajeevan, M. & Kshirsagar, S. R. Development of a high resolution daily gridded temperature data set (1969-2005) for the Indian region. *Atmos. Sci. Lett.* n/a-n/a (2009) doi:10.1002/asl.232.

59.    Junzeng, X., Qi, W., Shizhang, P. & Yanmei, Y. Error of Saturation Vapor Pressure Calculated by Different Formulas and Its Effect on Calculation of Reference Evapotranspiration in High Latitude Cold Region. *Procedia Eng.* **28**, 43–48 (2012).

60.    Hersbach, H. *et al.* The ERA5 global reanalysis. *Q. J. R. Meteorol. Soc.* **146**, 1999–2049 (2020).

61.    Hamed, K. H. & Ramachandra Rao, A. A modified Mann-Kendall trend test for autocorrelated data. *J. Hydrol.* **204**, 182–196 (1998).

62.    Granger, C. W. J. Investigating Causal Relations by Econometric Models and Cross-spectral Methods. *Econometrica* **37**, 424 (1969).

**Acknowledgments:**

RD thanks Dawn Emil Sebastian for initial support and help. The work is financially supported by Department of Science and Technology Swarnajayanti Fellowship Scheme, through project no. DST/ SJF/ E&ASA-01/2018-19; SB/SJF/2019-20/11, and Strategic Programs, Large Initiatives and Coordinated Action Enabler (SPLICE) and Climate Change Program through project no. DST/CCP/CoE/140/2018.

**Author contributions:**

SG conceived the idea and designed the problem. RD and SG performed the analysis. SG, RD, and RKC analyzed the results. SG and RD wrote the paper. RKC conducted a detailed review of the manuscript and did the necessary editing. AR and SK reviewed the manuscript.

**Competing interests:** The authors declare no competing interests.

# Supplementary Materials for

## Warming inhibits Increases in Vegetation Net Primary Productivity despite Greening in India

Ripan Das, Rajiv Kumar Chaturvedi, Adrija Roy, Subhankar Karmakar, Subimal Ghosh

**Affiliations**

[1]Interdisciplinary Program in Climate Studies, Indian Institute of Technology Bombay, Powai, Mumbai - 400 076, India

[2] Department of Humanities and Social Sciences, Birla Institute of Technology and Science-Goa Campus, Zuarinagar, India

[3]Department of Civil Engineering, Indian Institute of Technology Bombay, Powai, Mumbai - 400 076, India

[4]Environmental Science and Engineering Department, Indian Institute of Technology Bombay, Powai, Mumbai - 400 076, India

[*]Corresponding Author. Email: subimal@iitb.ac.in

**This PDF file includes:**

       Supplementary Note S1
       Supplementary Figures S1 to S21

**Supplementary Note 1:**

**Analytical solution of MODIS Leaf and Fine root Maintenance Respiration equation:**

We have looked into the equations for calculating MODIS NPP[1] and solved them analytically to investigate the effect of climate variables (mainly temperature) on net primary productivity. We have mainly focused on maintenance respiration since it is highly controlled by temperature.

MODIS NPP is calculated by the difference between GPP and Autotrophic Respiration (RA). The Autotrophic Respiration is a summation of Growth Respiration (RG) and Maintenance Respiration (RM). In the MODIS algorithm, RM = 0.25*NPP [1].

NPP = GPP – RA (RG+RM)      *Eq.* (1)

NPP = 0.8 (GPP – RM), When (GPP – RM) > 0 *Eq.* (2)

NPP = 0                      When (GPP – RM) < 0

The Maintenance Respiration term is the summation of leaf, fine root and deadwood respiration. The Leaf Maintenance Respiration (Leaf MR) is calculated using *Eq. 3*

$$Leaf\ MR = Leaf_{mass} \times Leaf_{mrbase} \times Q_{10}^{\frac{T_{avg}-20}{10}} \ Eq.(3)$$

$$Q_{10} = 3.22 - 0.046 \times T_{avg} \ \ Eq.(4)$$

$$Leaf\ MR = Leaf_{mass} \times Leaf_{mrbase} \times \left(3.22 - 0.046 \times T_{avg}\right)^{\frac{T_{avg}-20}{10}} \ Eq.(5)$$

We substitute , $3.22 = a$ , $0.046 = b$ and $Leaf_{mass} \times Leaf_{mrbase} = cl$ and differentiate *Eq. 5* with respect to T. *Eq.* 1-5 are taken from MODIS user guide [1].

$$\frac{d(Leaf\ MR)}{dT} = cl \times \left(a - b \times T_{avg}\right)^{\frac{T_{avg}-20}{10}} \times \left(\left(\frac{\ln{(a - b \times T_{avg})}}{10}\right) - \left(\frac{-b \times (T_{avg} - 20)}{10 \times (a - b \times T_{avg})}\right)\right) Eq.(6)$$

$$Froot\ MR = Fine_{rootmass} \times froot_{mrbase} \times Q_{10}^{\frac{T_{avg}-20}{10}} \ Eq.(7)$$

$$Froot\ MR = cr \times Q_{10}^{\frac{T_{avg}-20}{10}} Eq.(8)$$

Froot MR is the Fine root Maintenance Respiration. In the MODIS algorithm, $Q_{10}$ is constant value 2.0 for Froot[1]. We substitute $Fine_{rootmass} \times froot_{mrbase} = cr$ and differentiate $Eq.$ 8 with respect to T

$$\frac{d(Froot\ MR)}{dT} = cr \times Q_{10}^{\frac{T_{avg}-20}{10}} \times \frac{\ln(Q_{10})}{10} \ Eq.\ (9)$$

Putting the values of a, b in $Eq.$ 6 and $Q_{10}$ in $Eq.$ 9, we have plotted the temperature response of $\frac{1}{cl} \times \frac{d(Leaf\ MR)}{dT}$ and $\frac{1}{cr} \times \frac{d(Froot\ MR)}{dT}$ separately in the Supplementary Fig 18 (a) and (b) respectively.

We discovered that initially, Leaf MR rate increased with increasing temperature (Supplementary Fig 18a), but after a certain temperature limit, it began to fall. The temperature limit where Leaf MR rates fall will vary depending on biome type. We have taken $cl$ as constant since it is not a function of temperature. However, the Froot MR rate keeps on increasing with an increase in temperature (Fig S18b). The rate of increase will vary across biome types but will always be positive since $cr > 0$. We have taken $cr$ as constant since it is not a function of temperature.

For analyzing the collective effect of both leaf and froot MR rate changes with temperature, we have taken

$$\frac{d(Leaf\ and\ Froot\ MR)}{dT} = \frac{1}{cl} \times \frac{d(Leaf\ MR)}{dT} + \frac{1}{cr} \times \frac{d(Froot\ MR)}{dT} \ Eq.\ (10)$$

We have plotted the temperature response of $\frac{d(Leaf\ and\ Froot\ MR)}{dT}$ in the Supplementary Fig 18c. We can observe that, initially, leaf and respiration rates increase with an increase in temperature, but after a certain temperature, they remain stable. The temperature at which the rate of change will be stable will also vary across different biome type.

**Supplementary Figures:**

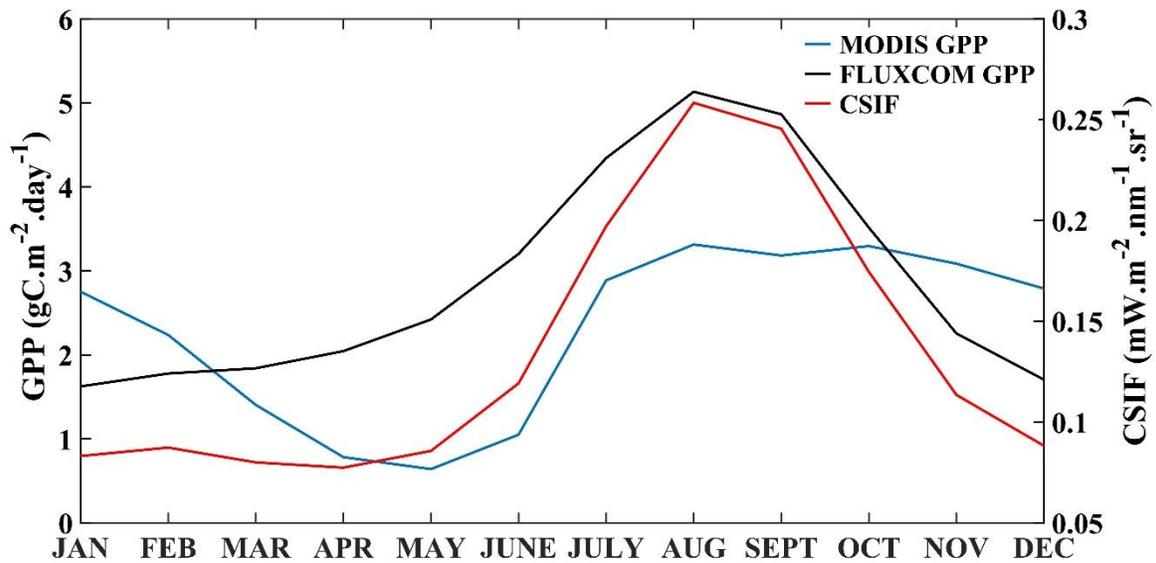

**Supplementary Figure 1. Climatology of Vegetation Productivity:** The climatology plot of MODIS GPP, FLUXCOM GPP and CSIF over India during the Period 2001-2018

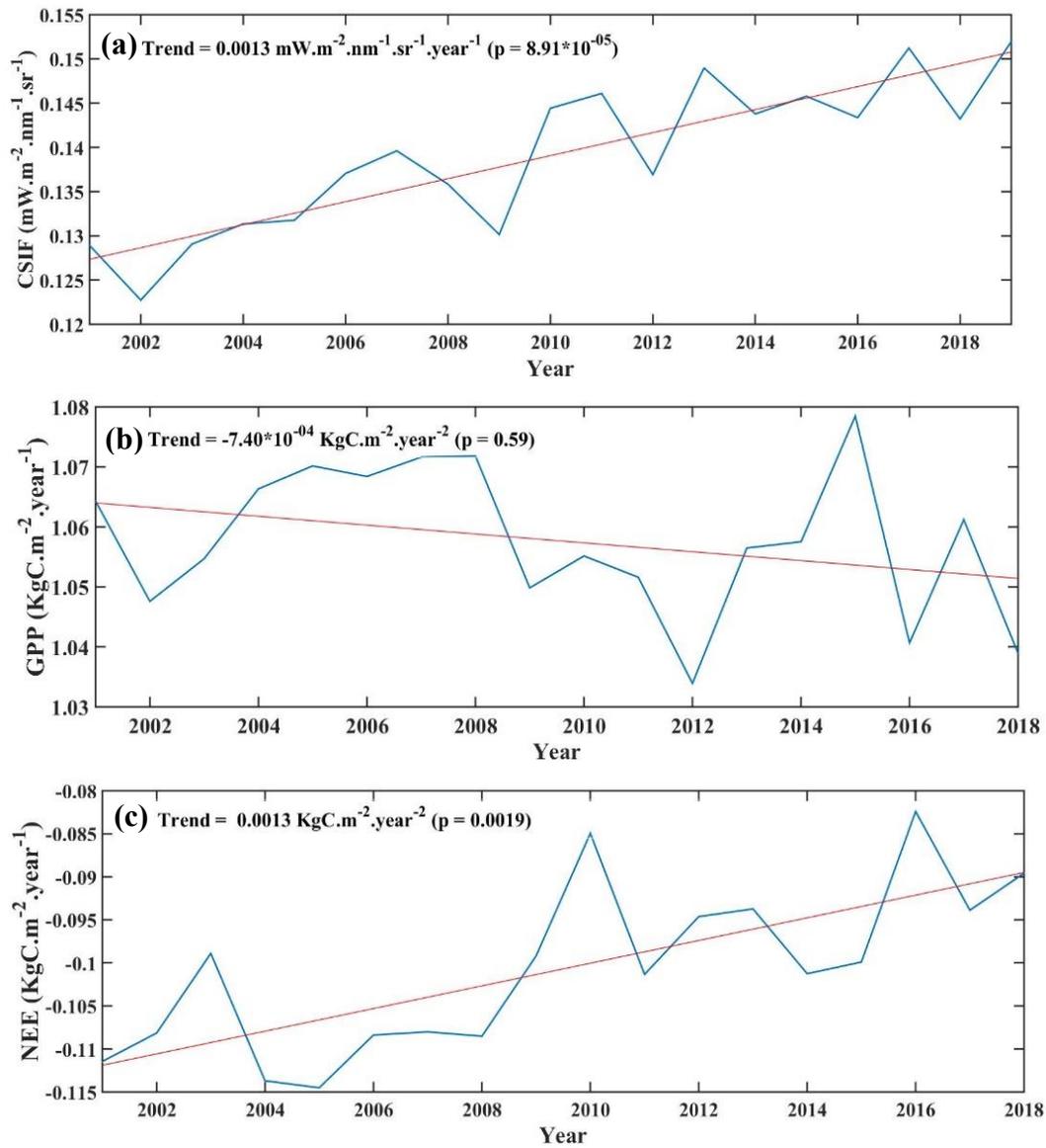

**Supplementary Figure 2. Time series plot of CSIF and FLUXCOM datasets:** The time series of CSIF (clear sky) (a), FLUXCOM GPP (b) and FLUXCOM NEE (c) over India. The CSIF dataset is during 2001-2019 and FLUXCOM datasets are during 2001-2018

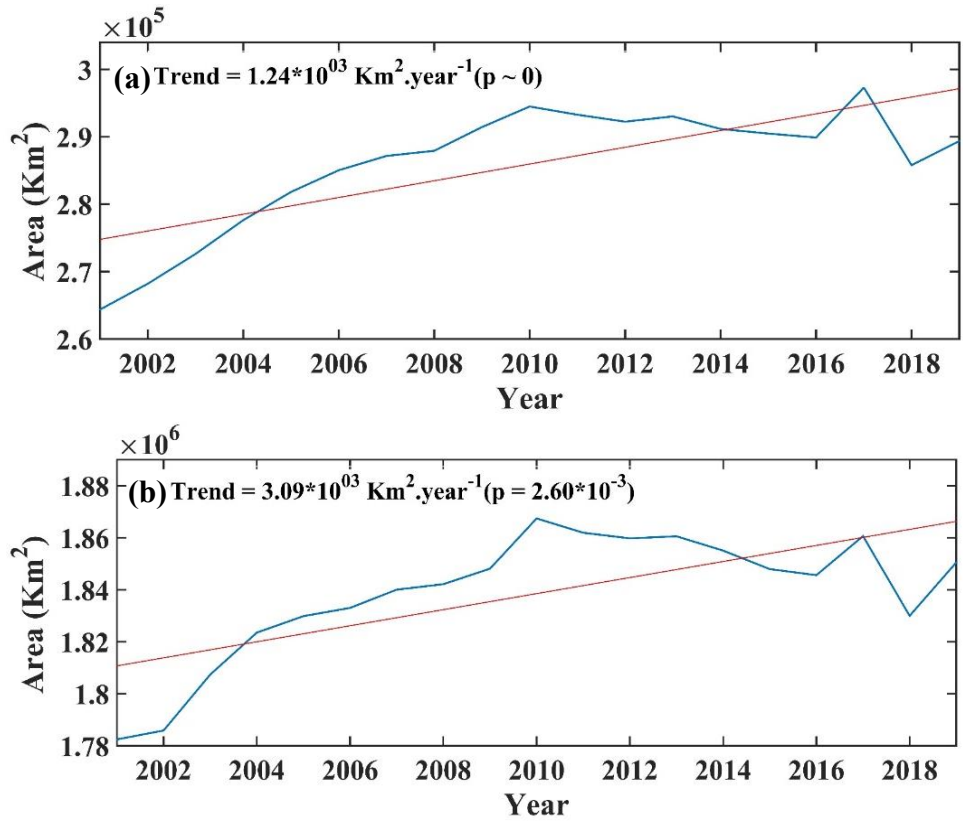

**Supplementary Figure 3. Greening in India:** The time series of Forest cover area (a) and Cropland area (b) over India from 2001 to 2019

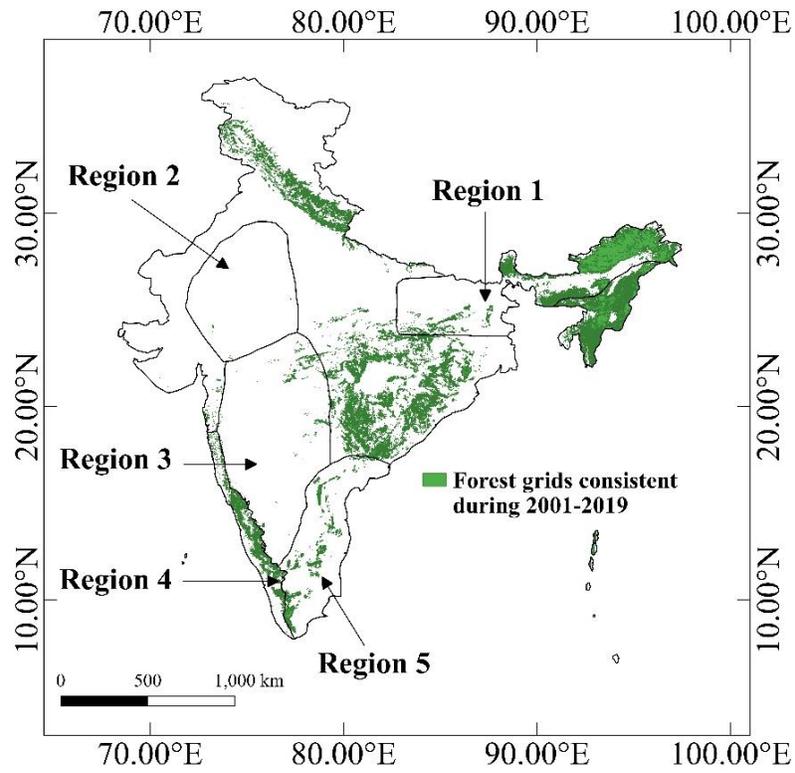

**Supplementary Figure 4. Consistent Forest cover:** Forest grids which are consistent during 2001-2019

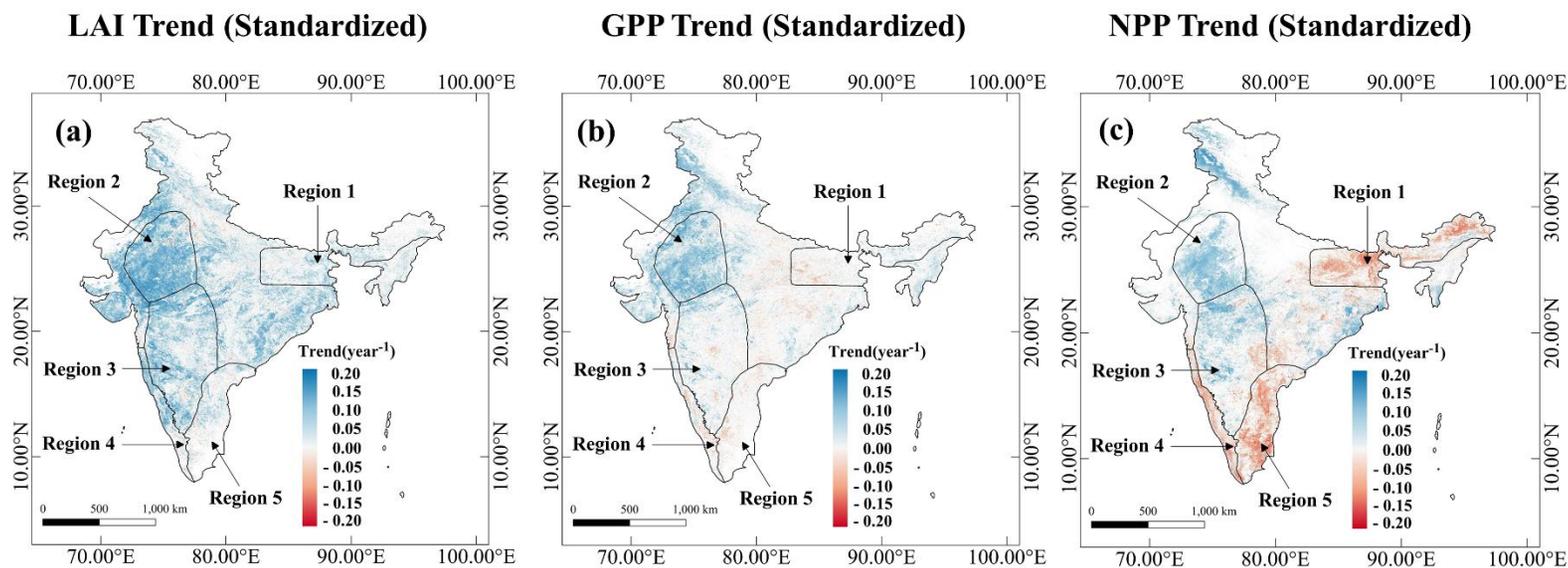

**Supplementary Figure 5. Regional Trends of Standardized MODIS datasets**: Trends of MODIS LAI (Standardized) (a), MODIS GPP (Standardized) (b), and MODIS NPP (Standardized) (c) at statistically significant level 0.1, Region 1: Northeast, Region 2: Northwest Arid, Region 3: Central Peninsular, Region 4: The Western Ghats, Region 5: East Coast Peninsular.

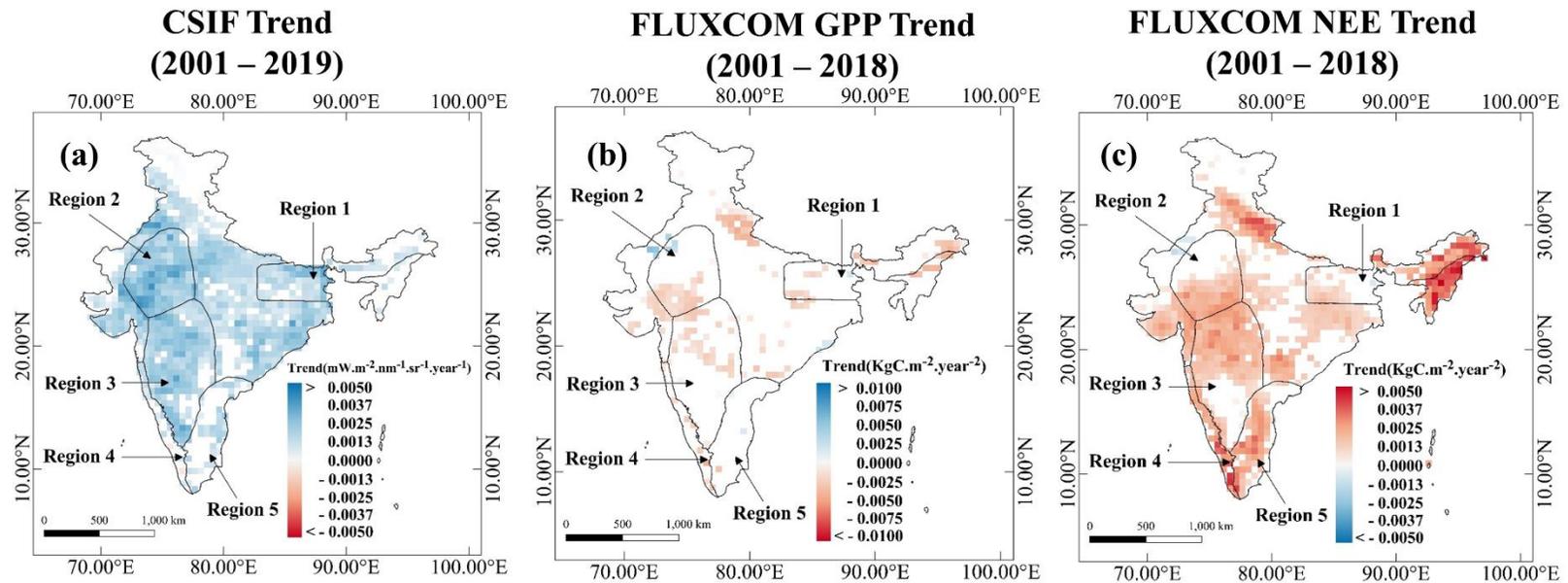

**Supplementary Figure 6. Regional Trends of CSIF and FLUXCOM datasets**: Trends of CSIF (a), FLUXCOM GPP (b), and FLUXCOM NEE (c) at statistically significant level 0.1, Region 1: Northeast, Region 2: Northwest Arid, Region 3: Central Peninsular, Region 4: The Western Ghats, Region 5: East Coast Peninsular.

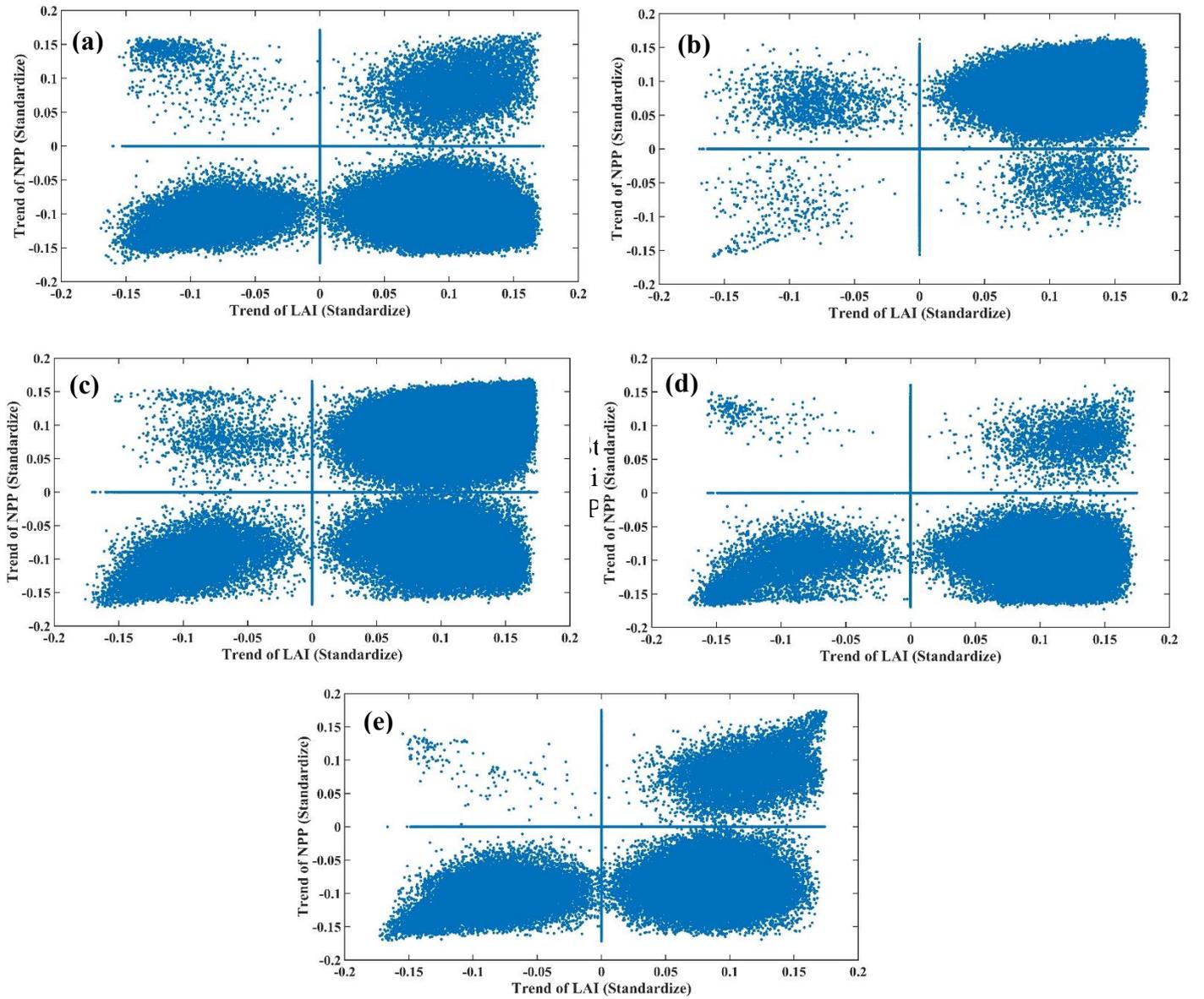

**Supplementary Figure 7. Divergence of Trend (Scatter):** Scatter plot of trends of standardize LAI and NPP for Region 1: Northeast (a), Region 2:  Northwest Arid (b), Region 3:  Central Peninsular (c), Region 4: The Western Ghats (d) and Region 5: East Coast Peninsular (e)

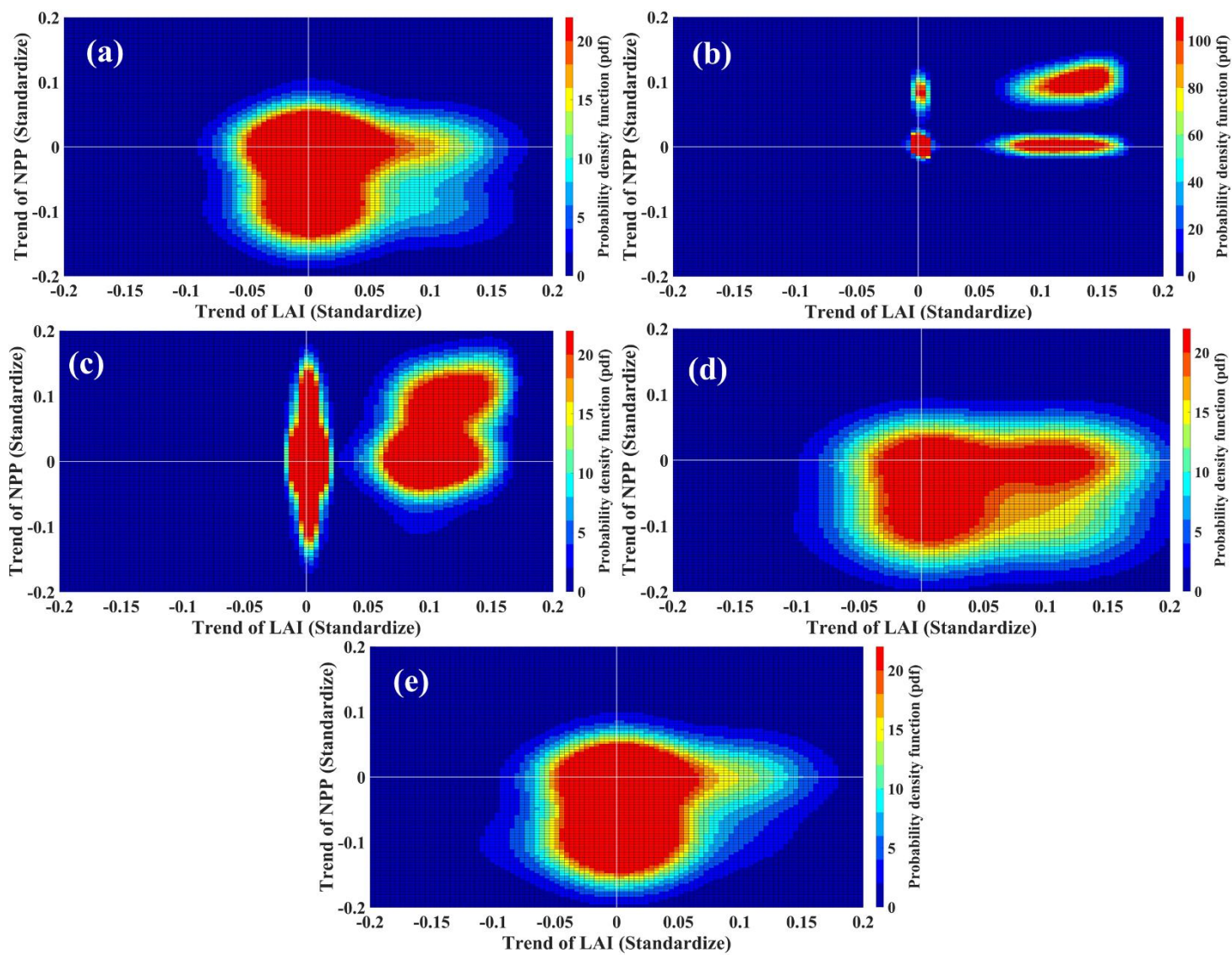

**Supplementary Figure 8. Divergence of Trend (density):** Probability density function (pdf) of the trends of standardized LAI and NPP trends using Bivariate kernel density estimate for Region 1: Northeast (a), Region 2: Northwest Arid (b), Region 3: Central Peninsular (c), Region 4: The Western Ghats (d) and Region 5: East Coast Peninsular (e)

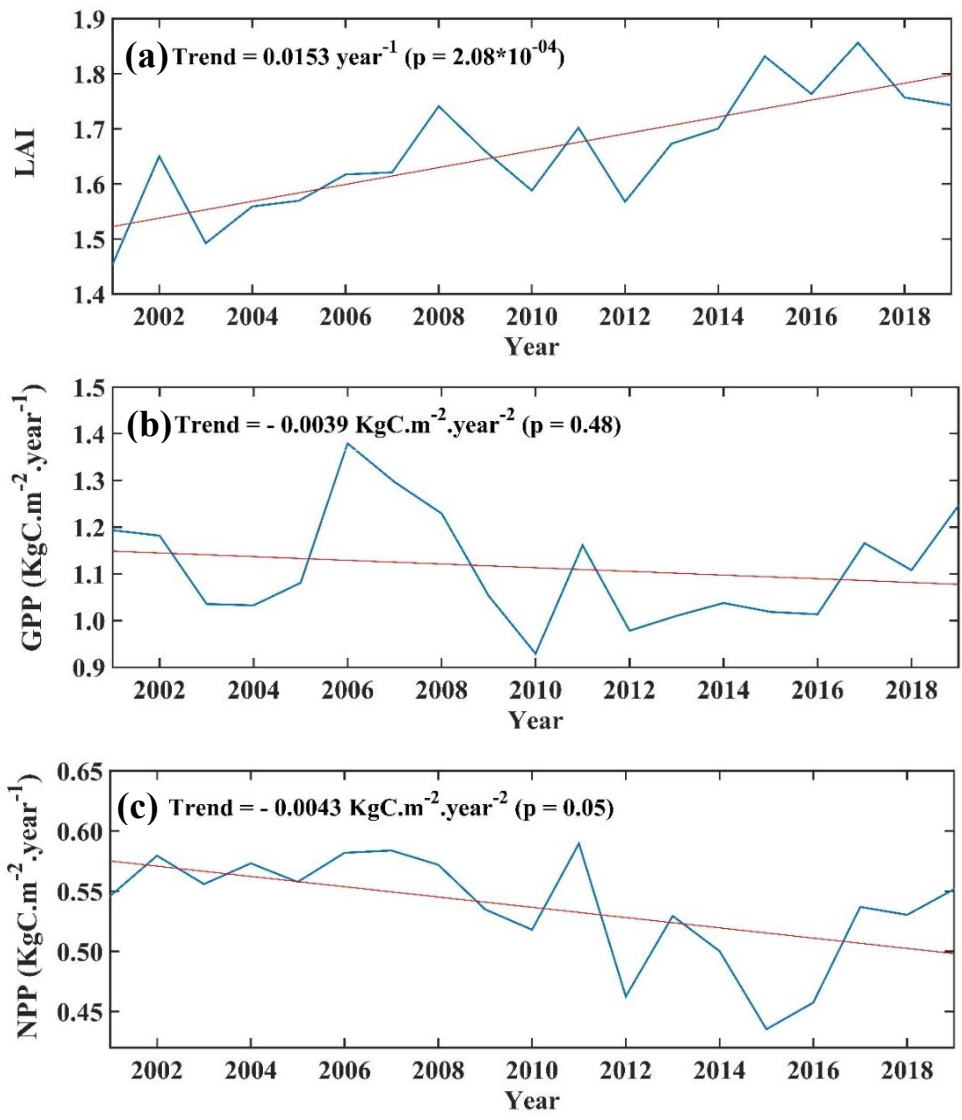

**Supplementary Figure 9. Regional Trend in the Northeast India:** The time series of LAI (a), GPP (b) and NPP (c) over Region 1: Northeast India during 2001-2019.

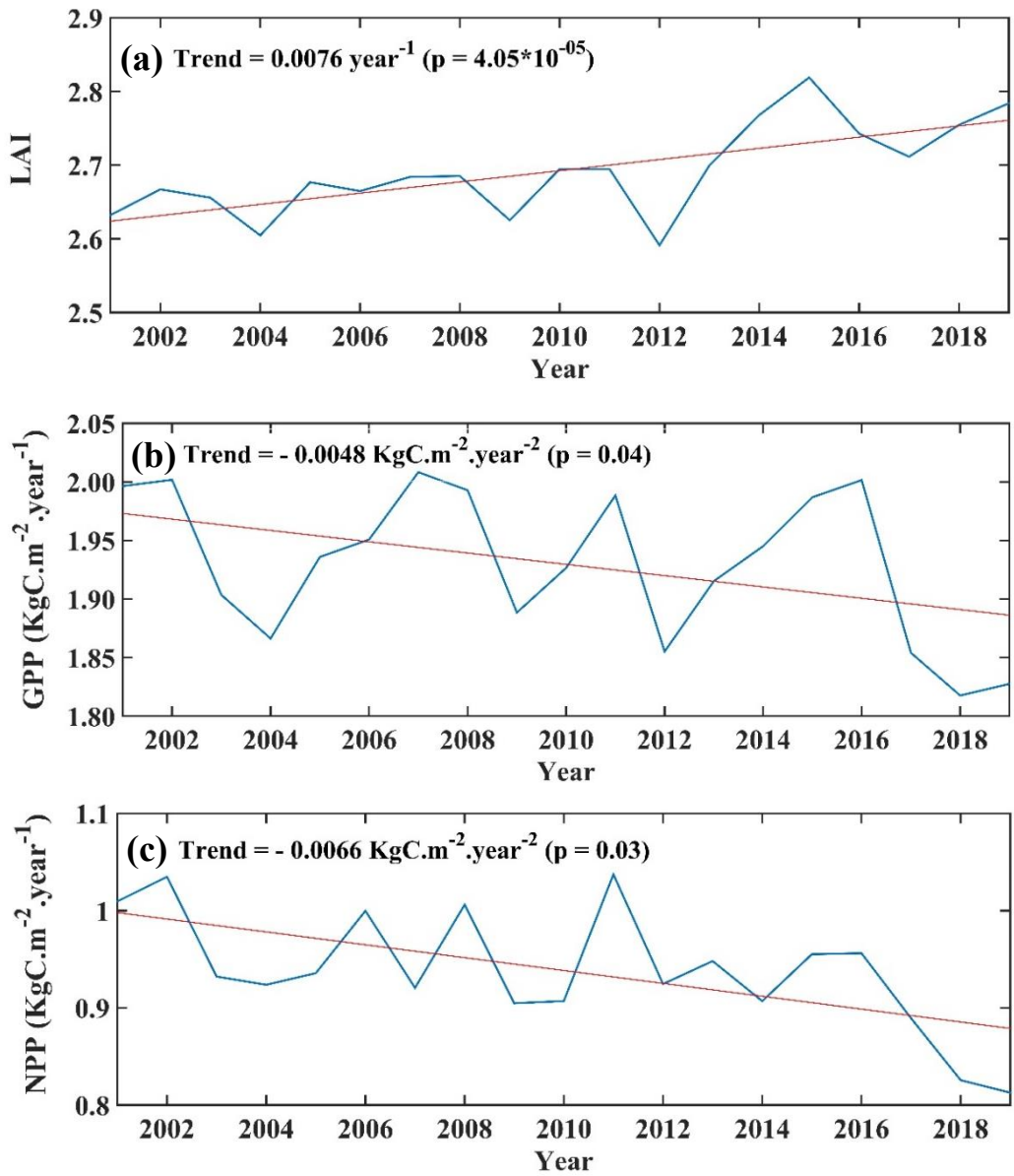

**Supplementary Figure 10. Regional Trend in the Western Ghats:** The time series of LAI (a), GPP (b) and NPP (c) over Region 4: The Western Ghats during 2001-2019.

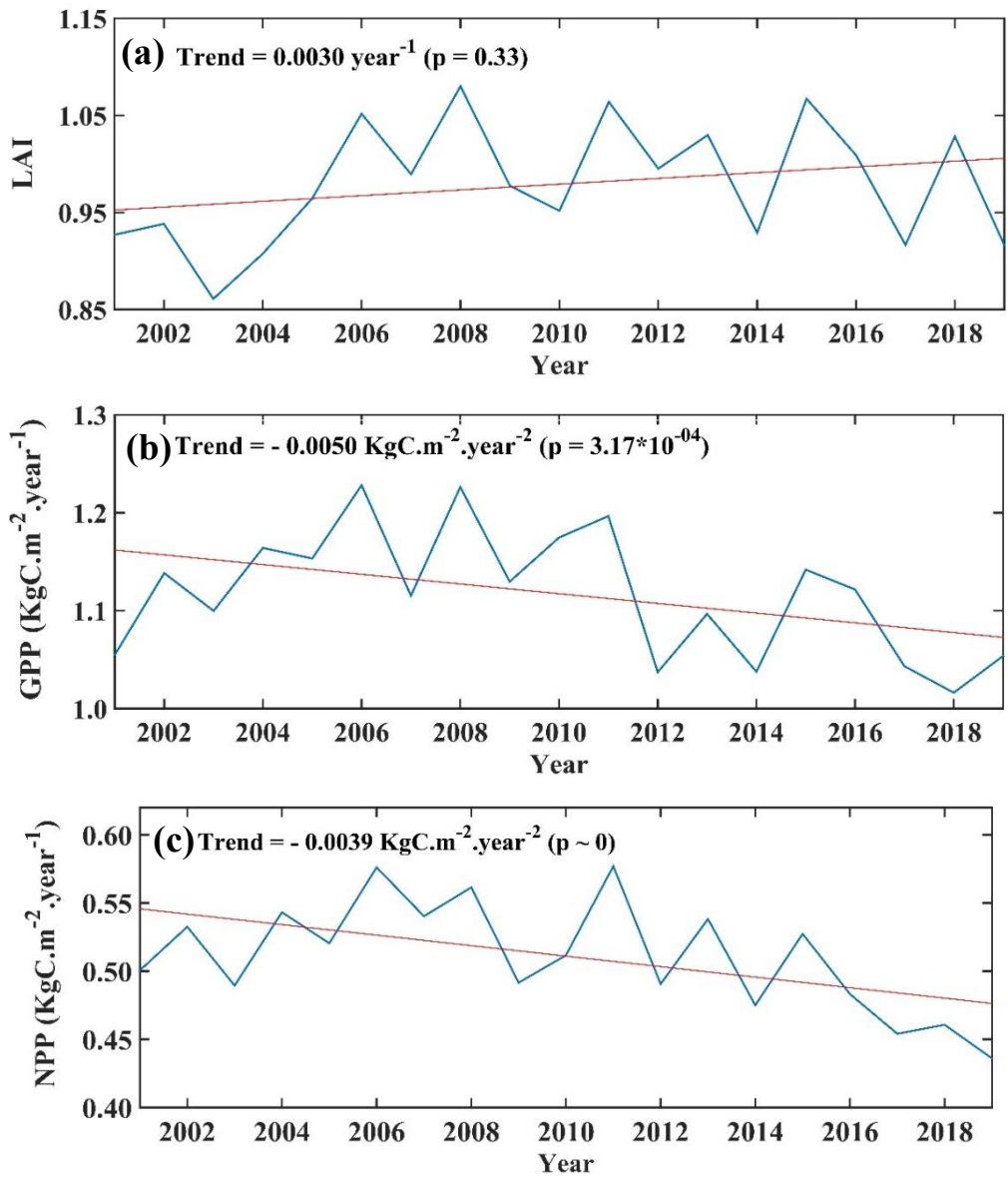

**Supplementary Figure 11. Regional Trend in the East Coast Peninsular India:** The time series of LAI (a), GPP (b) and NPP (c) over Region 5: East Coast Peninsular during 2001-2019.

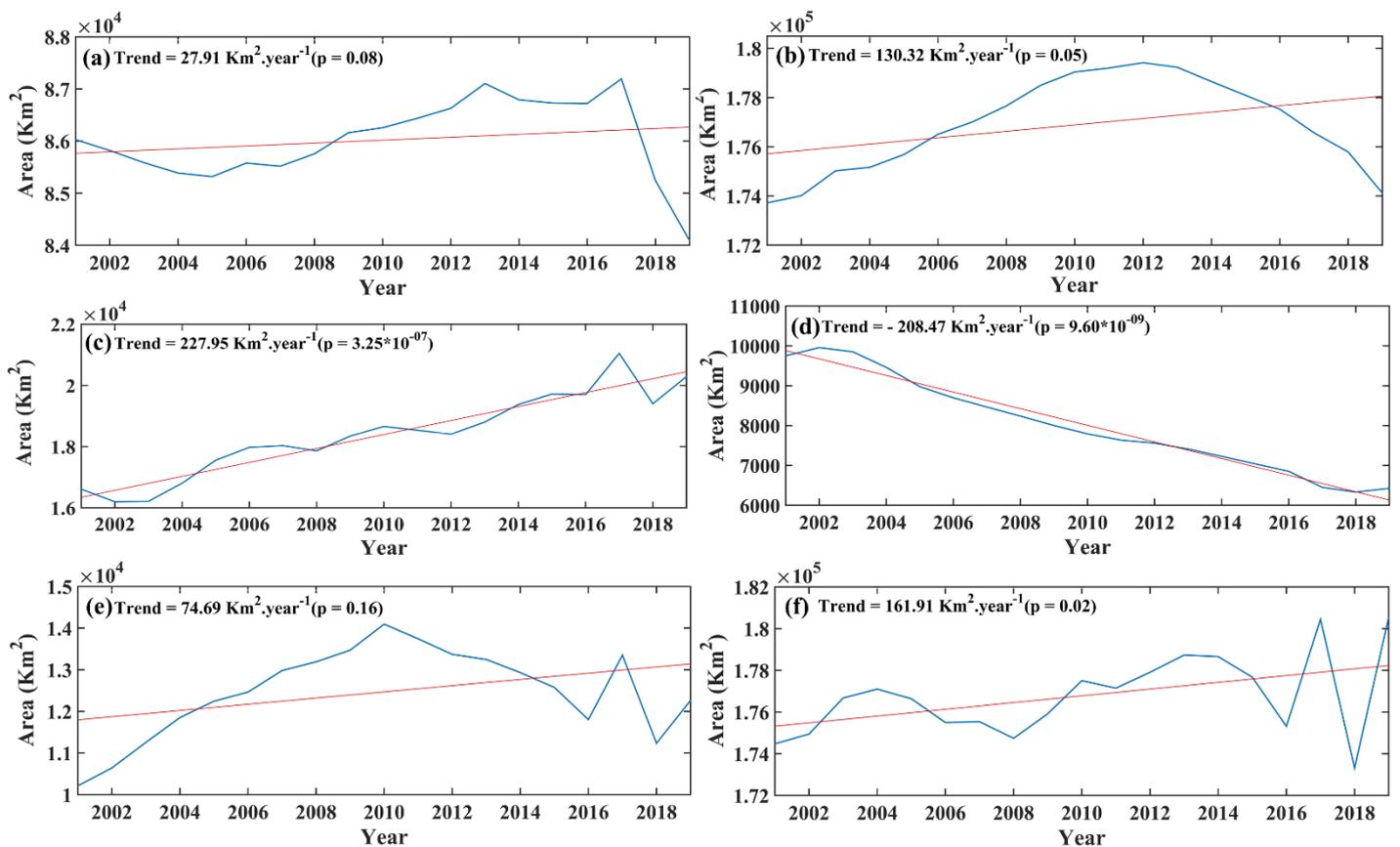

**Supplementary Figure 12. LULC Changes in the regions where LAI trends diverge from NPP trend:** Time series of Forest cover area Region 1: Northeast (a), Region 4: The Western Ghats (c) and Region 5: East Coast Peninsular (e). Time series of cropland area for Region 1: Northeast (b), Region 4: The Western Ghats (d) and Region 5: East Coast Peninsular (f).

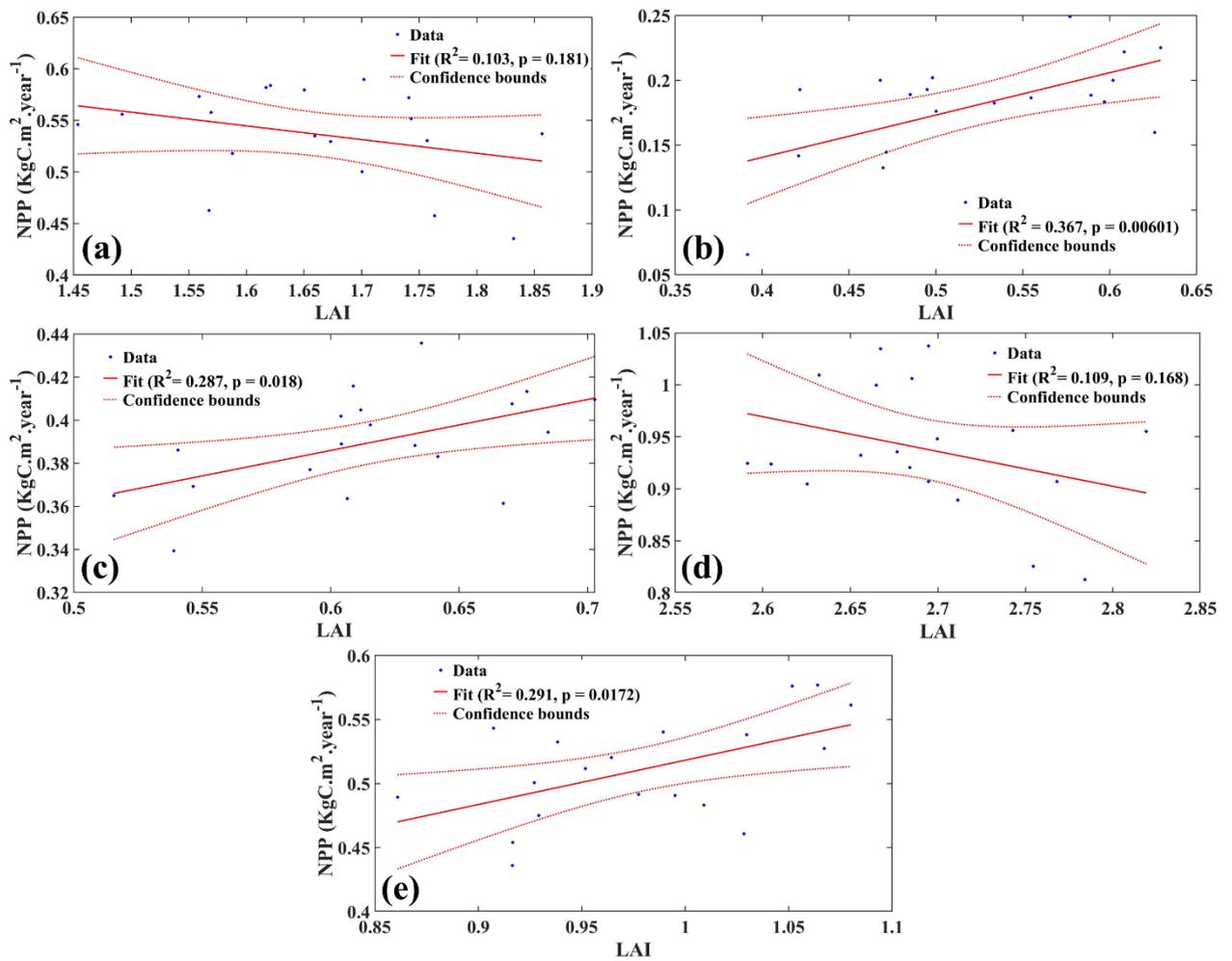

**Supplementary Figure 13. Relationship between LAI and NPP at Regional scale:** Liner Regression between annual LAI and NPP for Regions 1(a), 2(b), 3(c), 4(d), and 5 (e)

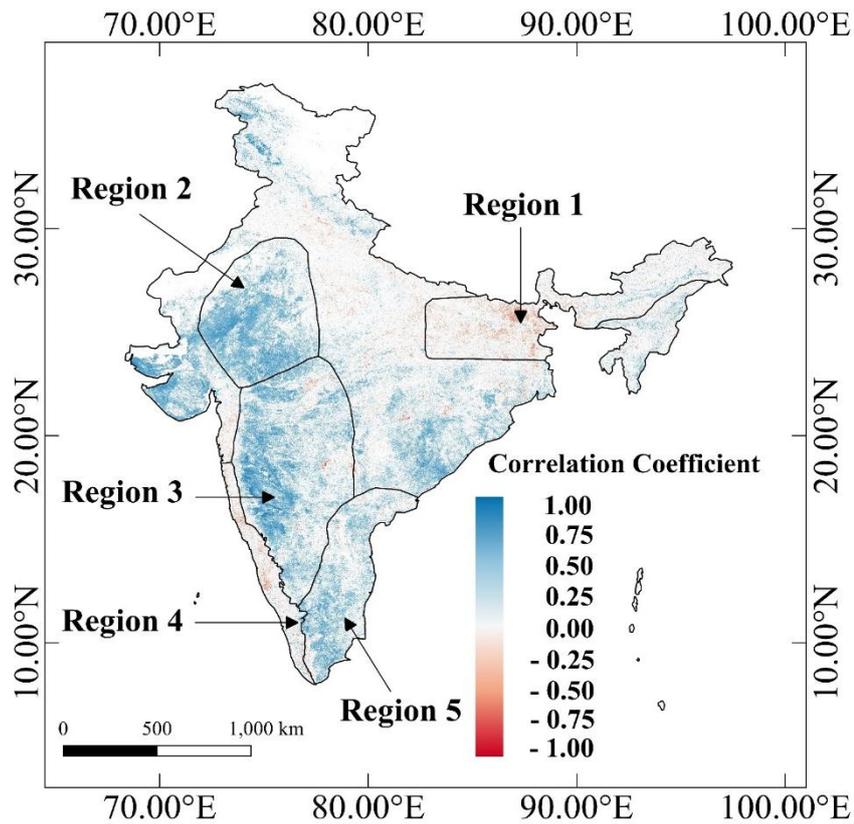

**Supplementary Figure 14. Correlation between LAI and NPP:** Pearson correlation coefficient map between NPP and Annual average LAI at statisticly significance level of 0.1

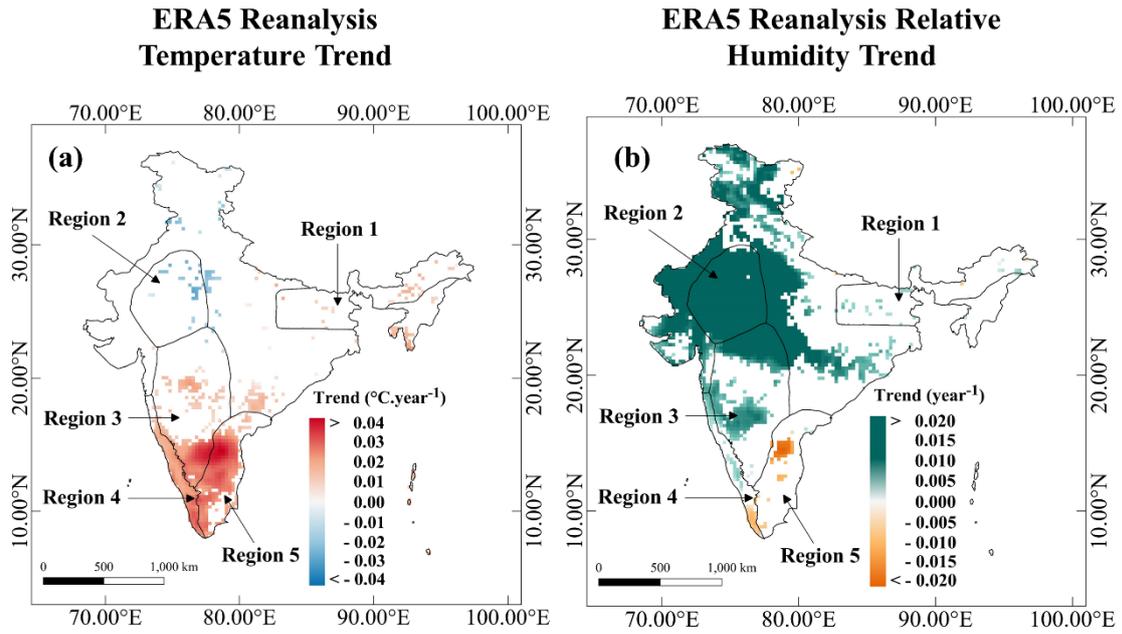

**Supplementary Figure 15. Relative Humidity and Temperature Trends:** Trend of ECMWF Relative Humidity (a) and ECMWF Temperature (b) in India for the period 2001-2019. The figures only show statistically significant trends at 0.1 level.

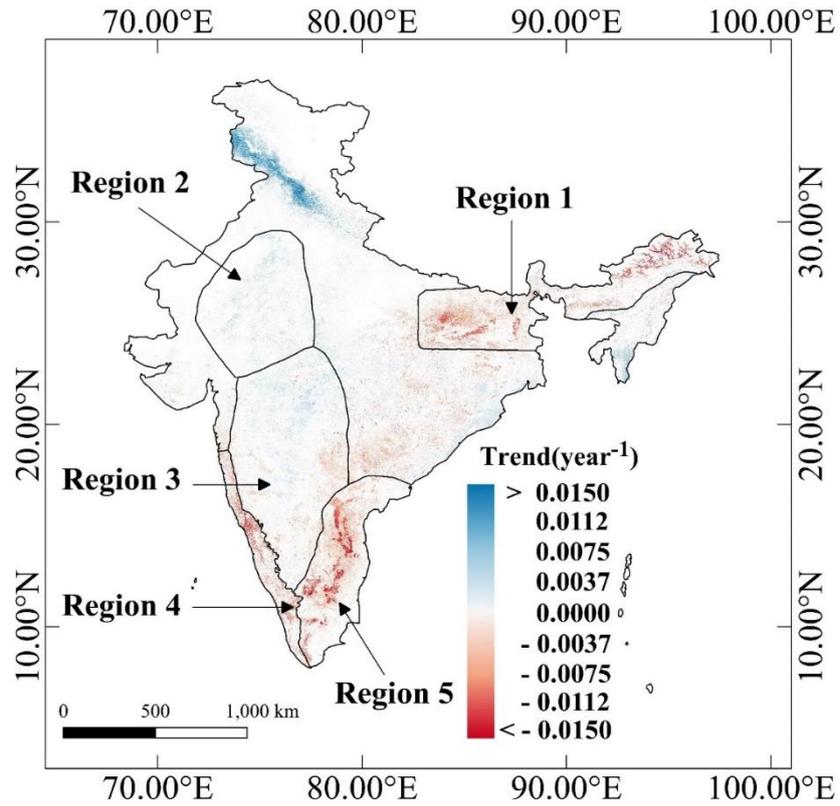

**Supplementary Figure 16. Residual Trend Map:** We linearly regressed NPP as function of LAI (*NPP = f (LAI)*) at each pixel during the period 2001-2019 over India and calculated the trend of the Residuals at a statistically significance level of 0.1.

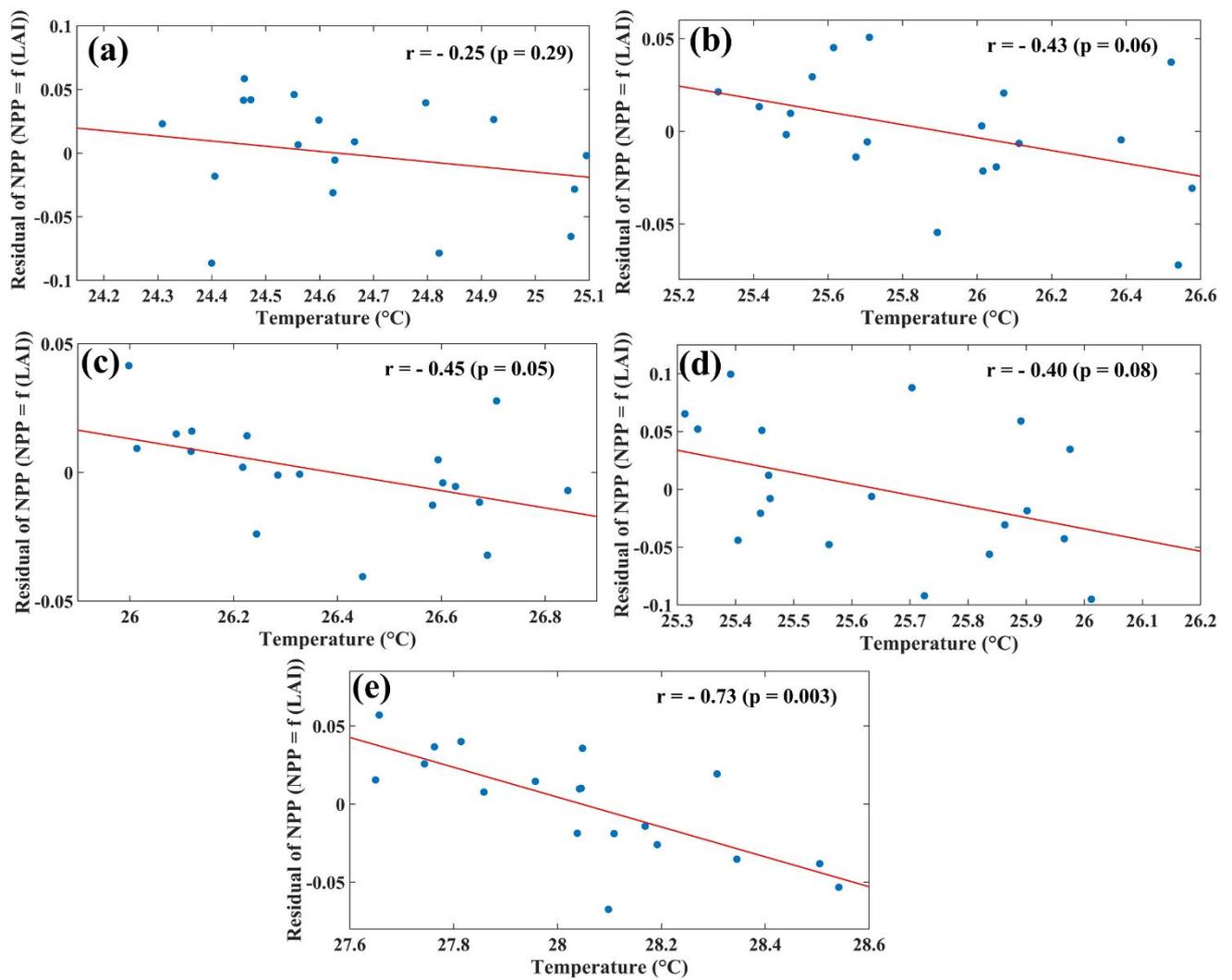

**Supplementary Figure 17. Impacts of Temperature on NPP:** Scatter plot between annual averaged temperature and residual of linearly regressed NPP as a function of annual LAI (NPP = f (LAI)) for Regions for 1(a), 2(b), 3(c), 4(d), 5 (e)

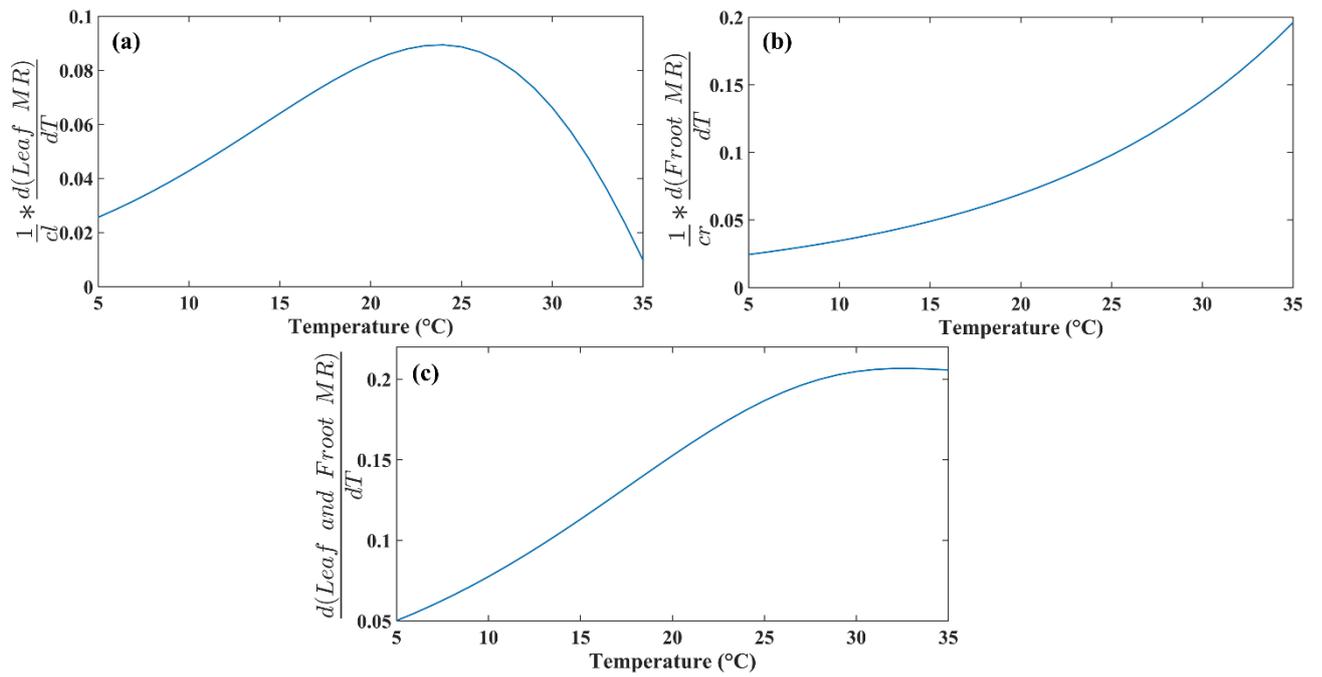

**Supplementary Figure 18. Temperature sensitivity of Leaf and Fine root Maintenance Respiration:** Sensitivity of $\frac{1}{cl} \times \frac{d(Leaf\ MR)}{dT}$ (a) , $\frac{1}{cr} \times \frac{d(Froot\ MR)}{dT}$ (b) and $\frac{d(Leaf\ and\ Froot\ MR)}{dT}$ (c) to temperature.

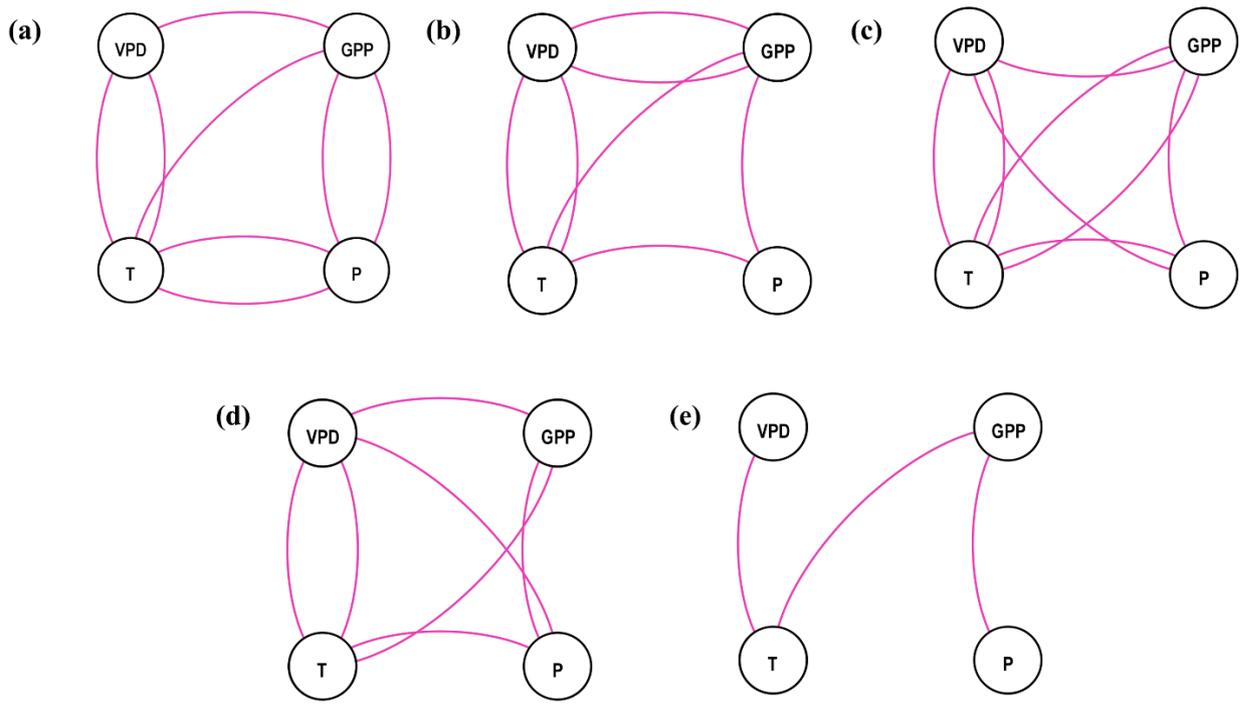

**Supplementary Figure 19. Climate impacting GPP in India:** The causal network derived from Granger Causality in the regions 1 (a), 2(b), 3(c), 4(d), 5 (e). The variables used are Precipitation (P), Temperature (T), Vapor Pressure Deficit (VPD), and Gross Primary Productivity (GPP). The causal links are taken from source to sink in a clockwise direction at statistically significance level 0.05

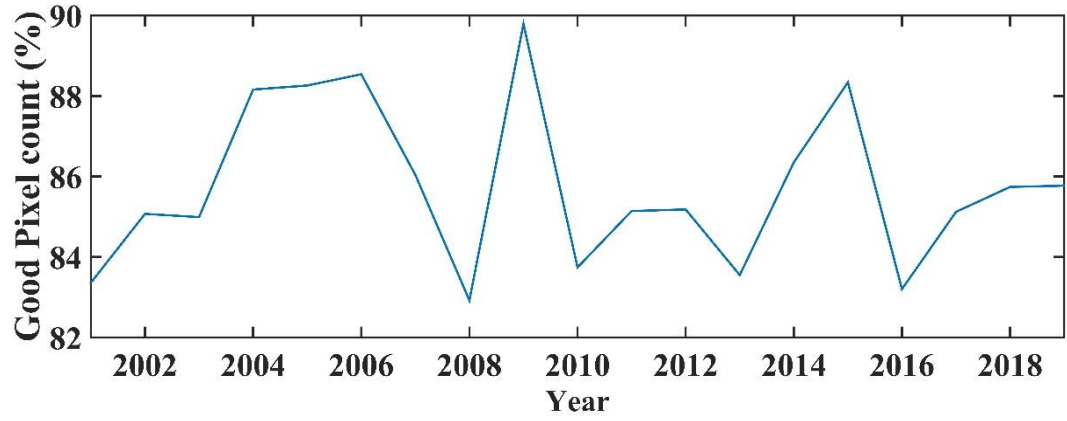

**Supplementary Figure 20. Unvaried Sample size:** Annual average good pixel (percentage) count over India during 2001 to 2019

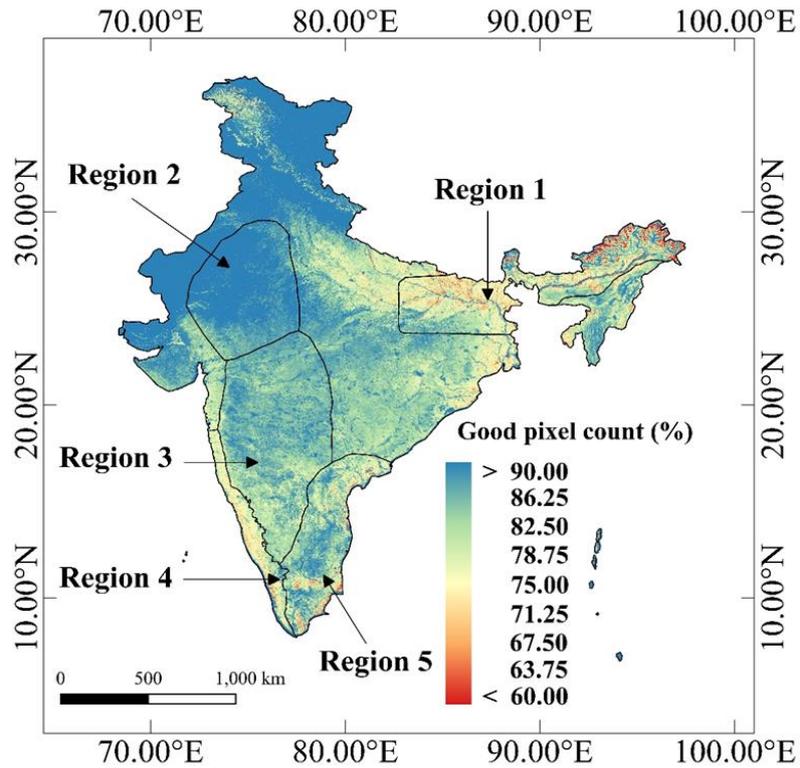

**Supplementary Figure 21. Satisfactory Pixel count:** Spatial map of good pixel count (percentage) from 2001 to 2019.

**References:**

1.  Running, S., Mu, Q., Zhao, M. & Moreno, A. MOD17A3HGF MODIS/Terra Net Primary Production Gap-Filled Yearly L4 Global 500 m SIN Grid V006 [Data set]. NASA EOSDIS Land Processes DAAC. (2019) doi:https://doi.org/10.5067/MODIS/MOD17A3HGF.006.